\documentclass[nohyper]{JHEP3}
\usepackage{amsmath,amssymb}
\usepackage{graphicx}
\usepackage{cite}
\usepackage{epsfig}
\usepackage{subfigure}

\def\gsim{\ \rlap{\raise 3pt \hbox{$>$}}{\lower 3pt \hbox{$\sim$}}\ }
 \def\lsim{\ \rlap{\raise 3pt \hbox{$<$}}{\lower 3pt \hbox{$\sim$}}\ }

\newcommand{\be}{\begin{equation}}
\newcommand{\ee}{\end{equation}}
\newcommand{\bea}{\begin{eqnarray}}
\newcommand{\eea}{\end{eqnarray}}
 \newcommand{\fh}{f_h^{eq}}
 \newcommand{\fht}{f_{\tilde{h}}^{eq}}
\newcommand{\fLeq}{f_{\ell}^{eq}} 
\newcommand{\fLteq}{f_{\widetilde{\ell}}^{eq}}

\title{Flavoured soft leptogenesis and natural values of the  B term}
\author{Chee Sheng Fong  \\ 
C.N. Yang Institute for Theoretical Physics\\
  State University of New York at Stony Brook\\
  Stony Brook, NY 11794-3840, USA.\\
  E-mail: \email{fong@insti.physics.sunysb.edu}}
\author{M.~C.~Gonzalez-Garcia\\
C.N. Yang Institute for Theoretical Physics\\
  State University of New York at Stony Brook\\
  Stony Brook, NY 11794-3840, USA {\rm and}  \\
  Instituci\'o Catalana de Recerca i Estudis Avan\c{c}ats (ICREA)\\
  Departament d'Estructura i Constituents de la Mat\`eria and ICC-UB\\
  Universitat de Barcelona, Diagonal 647, E-08028 Barcelona, Spain.\\
  E-mail: \email{concha@insti.physics.sunysb.edu}}
\author{Enrico Nardi\\
INFN, Laboratori Nazionali di Frascati\\
  C.P. 13, 100044 Frascati, Italy.\\
  E-mail: \email{enrico.nardi@lnf.infn.it}}
\author{J. Racker\\
  Departament d'Estructura i Constituents de la Mat\`eria and ICC-UB\\  
  Universitat de Barcelona, Diagonal 647, E-08028 Barcelona, Spain.\\
  E-mail: \email{racker@ecm.ub.es}}
\abstract{ We revisit flavour effects in soft leptogenesis relaxing
  the assumption of universality for the soft supersymmetry breaking
  terms. We find that with respect to the case in which the heavy
  sneutrinos decay with equal rates and equal CP asymmetries for all
  lepton flavours, hierarchical flavour configurations can enhance the
  efficiency by more than two orders of magnitude.  This translates in
  more than three orders of magnitude with respect to the one-flavour
  approximation.  We verify that lepton flavour equilibration effects
  related to off-diagonal soft slepton masses are ineffective for
  damping these large enhancements.  We show that soft leptogenesis
  can be successful for  unusual values of the relevant
  parameters, allowing for $B\sim {\cal O}({\rm TeV})$ and for values
  of the washout parameter up to $m_{\rm eff}/m_* \sim 5\times 10^{3}$.}
\keywords{Leptogenesis, Supersymmetry, Beyond Standard Model, Cosmology of Theories beyond the SM} 

\preprint{YITP-SB-10-12\\ICCUB-10-030}

\begin{document}

\section{Introduction}
The discovery of neutrino oscillations makes leptogenesis a very
attractive solution to the baryon asymmetry problem \cite{fy,leptoreview}.  
In the {\sl standard} type I seesaw  framework \cite{ss}, the   
singlet heavy neutrinos have  lepton number violating 
Majorana masses and when decay out of equilibrium produce 
dynamically a lepton asymmetry which  is partially converted into a
baryon asymmetry due to fast sphaleron processes.

For a hierarchical spectrum of the $SU(2)$ singlets Majorana neutrinos,
successful leptogenesis requires generically quite heavy singlet
neutrino masses~\cite{di}, of order $M>2.4 (0.4)\times 10^9$~GeV for
vanishing (thermal) initial neutrino densities~\cite{di,Mbound}
(although flavour effects \cite{flavour1,flavour2,db2,oscar} and/or
extended scenarios \cite{db1,ma} may affect this limit).  Low-energy
supersymmetry can be invoked to naturally stabilize the hierarchy
between this new scale and the electroweak one. This, however,
introduces a certain conflict between the gravitino bound on the
reheat temperature and the thermal production of the heavy singlets 
neutrinos \cite{gravi}.  A way out of this conflict is provided by
resonant leptogenesis~\cite{PU}.  In this scenario, the heavy Majorana  
neutrinos are nearly degenerate in mass which makes the self energy
contributions to the CP asymmetries resonantly enhanced, thus allowing
for successful leptogenesis at much lower temperatures.

Once supersymmetry has been introduced, leptogenesis is induced also
in singlet sneutrino decays.  If supersymmetry is not broken, the
order of magnitude of the asymmetry and the basic mechanism are the
same as in the non-supersymmetric case. However, as shown in
Refs.\cite{soft1,soft2}, supersymmetry-breaking terms can induce
effects which are essentially different from the neutrino ones.  In
brief, soft supersymmetry-breaking terms involving the singlet
sneutrinos remove the mass degeneracy between the two real sneutrino
states of a single neutrino generation, and provide new sources of
lepton number and CP violation.  In this case, as for the case of
resonant leptogenesis, it is the sneutrino self-energy contributions
to the CP asymmetries which are resonantly enhanced.  As a
consequence, the mixing between the sneutrino states can generate a
sizable CP asymmetry in their decays.  This scenario was termed ``soft
leptogenesis''~\cite{soft2}.

Altogether it was found that the asymmetry is large for a Majorana
neutrino mass scale relatively low, in the range $10^{5}-10^{8}$ GeV.
A sizable part of this range lies below the reheat temperature
limits, what solves the cosmological gravitino problem.  However, in
order to generate enough asymmetry the lepton-violating soft bilinear
coupling, $B$, responsible for the sneutrino mass splitting, has to be
unconventionally small~\cite{soft1,soft2,oursoft,ourqbe}
\footnote{Flavour effects~\cite{ourgaugino} and extended
  scenarios~\cite{ourinvsoft,softothers} may alleviate the
  unconventionally-small-$B$ problem.}.

In Refs.~\cite{soft3,ourgaugino} the possibility of soft leptogenesis
generated by CP violation in the decay of the heavy sneutrinos, and in
the interference of mixing and decay was considered.  These new
sources of CP violation (the so called ``new ways to soft
leptogenesis''~\cite{soft3}) are induced by vertex corrections due to
gaugino soft supersymmetry-breaking masses.  In all these processes,
at first order in soft breaking terms and at $T=0$, the CP asymmetries
for the decays into fermions and bosons are equal in magnitude and
opposite in sign~\cite{soft1,soft2,ourgaugino}.  Therefore, assuming
equality in the time evolutions of the fermion and scalar lepton
asymmetries (that at sufficiently low temperatures is certainly
correct because of super-equilibration of the particle-sparticle
chemical potentials~\cite{superequilibration}) in the $T=0$ limit an
exact cancellation occurs between the lepton asymmetry produced in the
fermionic and bosonic channels~\cite{soft1,soft2,ourgaugino}.  Thermal
effects, thus, play a fundamental role in soft leptogenesis:
final-state Fermi blocking and Bose stimulation as well as effective
masses for the particle excitations in the plasma break supersymmetry
and effectively spoil the cancellation.

In Refs.~\cite{soft1,soft2,soft3} soft leptogenesis was addressed
within the `one-flavour' approximation. This one-flavour approximation
is rigorously correct only when the interactions mediated by charged
lepton Yukawa couplings are out of equilibrium.  This is not the case
in soft leptogenesis since, as mentioned above, successful
leptogenesis in this scenario requires a relatively low mass scale for
the singlet neutrinos.  Thus the characteristic temperature is such
that the rates of processes mediated by the $\tau$ and $\mu$ Yukawa
couplings are not negligible, implying that the effects of lepton
flavours must be taken into account.  The impact of flavour in thermal
leptogenesis in the context of standard see-saw leptogenesis has been
investigated in great detail in several
papers~\cite{flavour1,flavour2,db1,oscar,%
  flavourothers,barbieri,PU,riottoqbefla,riottosc}. In general, the
result of including flavour effects is that the produced baryon
asymmetry can get considerably  enhanced, and some of the constraints on
the required value of the heavy Majorana neutrino and sneutrino masses
can be relaxed.  The effects of spectator processes, that are fast
processes that do not violate lepton number but that can still have an
impact on lepton asymmetry production, was analyzed
in~\cite{spectator1,spectator2}.  It was found that within the
standard leptogenesis scenario the size of the related corrections is
at most of ${\cal O}(1)$, and less important than effects related to
the lepton flavours.  A quite general characteristic of models for new
physics, like for example the MSSM, is the presence of new sources of
lepton flavour violation, that are not suppressed by the light neutrino
masses. As has been highlighted in ref.~\cite{lfe}, at sufficiently  
low temperatures  the related new effects can give rise to  
lepton flavour equilibration (LFE), and when this occurs all dynamical 
flavour  effects get effectively damped.

Ref.~\cite{oursoft} introduced flavour and spectator effects in the
soft leptogenesis scenario under the restrictive assumption of
universal trilinear couplings. The authors found that within that
context, these effects could enhance the efficiency by ${\cal O }(30)$.

In this work we revisit the impact of flavour in soft leptogenesis.
In particular we extend the analysis of Refs.~\cite{oursoft,lfe} by
relaxing the assumption of universal trilinear couplings.  We find
that under these conditions flavour effects can play an even more
important role, enhancing the leptogenesis efficiency by more than 
three orders of magnitude with respect to the one-flavour approximation.
Given the importance that flavour effects can acquire in soft
leptogenesis with non-universal soft supersymmetry breaking terms, we
also quantify the LFE effects associated with off-diagonal soft
breaking masses for the scalar lepton doublets. We find that in most
part of the supersymmetry (SUSY) parameter space that is relevant for
soft leptogenesis, the large flavour enhancements can survive LFE
effects.

The outline of the paper is as follows.  Section~\ref{sec:lag}
summarizes the soft leptogenesis scenario for the general flavour
structure of the relevant trilinear couplings, and presents the
corresponding CP violation asymmetries.  In Sec.~\ref{sec:lfe} we
review the flavour changing processes associated with off-diagonal
soft breaking masses for the scalar lepton doublets, and we discuss
the temperature regime in which LFE becomes relevant.  In
Sec.~\ref{sec:results} we present the relevant Boltzmann Equations
(BE) that describe the production of the lepton asymmetry in this
scenario.  To quantify the achievable flavour enhancements as well as
the possible impact of LFE effects, we then solve the BE for different
flavour configurations and in different temperature regimes.
In Sec.~\ref{sec:disc} we discuss the
possible connection with observable lepton flavour violation phenomena
at low energies.  Finally, in Sec.~\ref{sec:concl} we summarize our
results and draw the conclusions.

\section{Soft Leptogenesis Lagrangian and CP Asymmetries}
\label{sec:lag}

The supersymmetric see-saw model can be described by the superpotential:
\begin{equation}
W=\frac{1}{2}M_{ij}N_{i}N_{j}+Y_{ik}
\epsilon_{\alpha\beta}N_{i}L_{k}^{\alpha}H^{\beta},
\label{eq:superpotential}
\end{equation}
where $i,j$ are the generation indices of heavy Majorana
`right-handed' (RH) neutrinos, $k=1,2,3$ are the lepton flavour
indices, and $N_{i}$, $L_{k}$, $H$ are the chiral superfields for the
RH neutrinos, the left-handed (LH) lepton doublets and the Higgs
doublets, with $\epsilon_{\alpha\beta}=-\epsilon_{\beta\alpha}$ and
$\epsilon_{12}=+1$.

The relevant soft supersymmetry breaking terms involving the RH
sneutrinos $\widetilde{N_{i}}$ and $SU(2)$ gauginos
$\widetilde{\lambda}_{2}^{a}$ are given by \footnote{The effect of
$U(1)$ gauginos can be included in similar form.}
\begin{eqnarray} 
\mathcal{L}_{soft} & = 
& -\left(A Z_{ik}\epsilon_{\alpha\beta}\widetilde{N}_{i}
\tilde{\ell}_{k}^{\alpha}h_2^{\beta}+\frac{1}{2}B_{ij}
M_{ij}\widetilde{N}_{i}\widetilde{N}_{j}
+\frac{1}{2}m_{2}\overline{\tilde{\lambda}}_{2}^{a}
P_{L}\tilde{\lambda}_{2}^{a} +\mbox{h.c.}\right)\; .
\label{eq:soft_terms}
\end{eqnarray}
The sneutrino and anti-sneutrino states mix, giving rise to the mass
eigenstates:
\begin{eqnarray}
\widetilde{N}_{+i} & = &
\frac{1}{\sqrt{2}}(e^{i\Phi/2}\widetilde{N}_{i}+e^{-i\Phi/2}
\widetilde{N}_{i}^{*}),\nonumber
\\ \widetilde{N}_{-i} & = &
\frac{-i}{\sqrt{2}}(e^{i\Phi/2}
\widetilde{N}_{i}-e^{-i\Phi/2}\widetilde{N}_{i}^{*}),
\label{eq:mass_eigenstates}
\end{eqnarray}
where $\Phi\equiv\arg(BM)$, that correspond to the mass eigenvalues
\begin{eqnarray}
M_{ii\pm}^{2} & = & M_{ii}^{2} \pm|B_{ii}M_{ii}|.
\label{eq:mass_eigenvalues}
\end{eqnarray}

The Lagrangian for the interactions involving the RH sneutrinos
$\widetilde{N}_{\pm i}$, the RH neutrinos $N_{i}$ and the $SU(2)$
gauginos $\tilde{\lambda}_{2}$, with the (s)leptons and the
Higgs(inos) can be written as:
\begin{eqnarray}
\mathcal{L}_{int}&=&
-\epsilon_{\alpha\beta}\left\{\frac{1}{\sqrt{2}} 
\widetilde{N}_{+i}\left[Y_{ik}\overline{\widetilde{h}}^{\beta}P_{L}\ell_{k}^{\alpha} +(A Z_{ik} + M_{i} Y_{ik})
\widetilde{\ell}_{k}^{\alpha}h_2^{\beta}\right] \right.
\nonumber \\
&& \left.
+\frac{i}{\sqrt{2}}\widetilde{N}_{-i}\left[Y_{ik}\overline{\widetilde{h}}^{\beta}P_{L}\ell_{_{k}}^{\alpha}
+(A Z_{ik} -M_{i} Y_{ik})\widetilde{\ell}_{k}^{\alpha}h_2^{\beta}\right]
+Y_{ik} \overline{\widetilde{h}}^{\beta}P_{L}N_{i}\widetilde{\ell}_{k}^{\alpha}
+ Y_{ik} \overline{N}_i P_L\ell_k^{\alpha}h_2^{\beta}\right\}
\nonumber \\
&& -g_{2}\left(
\overline{\widetilde{\lambda}}_{2}^{\pm}P_{L}
(\sigma_{1})_{\alpha\beta}\ell_{k}^{\alpha}\widetilde{\ell}_{k}^{\beta*}
-\frac{1}{\sqrt{2}}\overline{\widetilde{\lambda}}_{2}^{0}P_{L}
(\sigma_{3})_{\alpha\beta}\ell_{k}^{\alpha}\tilde{\ell}_{k}^{\beta*}
\right.\nonumber \\
 &  & \left.
+\overline{\widetilde{h}}^{\alpha}P_{L}(\sigma_{1})_{\alpha\beta}
\widetilde{\lambda}_{2}^{\pm}h_2^{\beta*}-\frac{1}{\sqrt{2}}
\overline{\widetilde{h}}^{\alpha}P_{L}(\sigma_{3})_{\alpha\beta}
\widetilde{\lambda}_{2}^{0}h_2^{\beta*}\right)
+\mbox{h.c.}\; ,
\label{eq:mass_basis}
\end{eqnarray}
where
$\ell_{k}^{T}=\left(\nu_{k},\ell_{k}^{-}\right)$,
$\widetilde{\ell}_{k}^{T}=
\left(\widetilde{\nu}_{k},\widetilde{\ell}_{k}^{-}\right)$ are the
lepton and slepton doublets, and
$h_2^{T}=\left(h_2^{+},h_2^{0}\right)$,
$\widetilde{h}^{T}=\left(\widetilde{h}^{-},\widetilde{h}^{0}\right)$
are the Higgs and Higgsino doublets.  $\widetilde{\lambda}_{2}^{\pm}$
denotes $\widetilde{\lambda}_{2}^{+}$ for $(\alpha\beta)=(01)$ and
$\widetilde{\lambda}_{2}^{-}$ for $(\alpha\beta)=(10)$ with
$\sigma_{1,3}$ being the Pauli matrices, and $P_{L,R}$ are
respectively the left and right projection operator.

All the parameters appearing in the superpotential
Eq.~\eqref{eq:superpotential} and in the Lagrangian
Eq.~\eqref{eq:soft_terms} (and equivalently in the first two lines of
Eq.~\eqref{eq:mass_basis}) are in principle complex quantities.
However, superfield phase redefinition allows to remove several
complex phases. Here for simplicity, we will concentrate on soft
leptogenesis arising from a single sneutrino generation $i=1$ and in
what follows we will drop that index ($Y_k\equiv Y_{1k},\, Z_{k}\equiv
Z_{1k},\, B=B_{11},$ etc.). Thus we will be only interested in the
physical phases involving the sneutrinos of the first generation.
After superfield phase rotations, the relevant Lagrangian terms 
restricted to $i=1$ are characterized by only  
four independent physical phases, that are 
\begin{eqnarray}
\label{eq:CPphase1}
\phi_{Ak}&=&{\rm arg}(Z_{k} Y^*_k  A B^*), \qquad (k=1,2,3) \\
\phi_{g}&=&\frac{1}{2}{\rm arg}(B m_2^*), 
\label{eq:CPphase2}
\end{eqnarray}
which we choose to assign respectively to the slepton-Higgs-sneutrino
trilinear soft breaking terms, and to the gaugino coupling 
operators respectively.
In what follows we will keep track of these physical phases explicitly
and, differently from the convention used in
Eqs.~\eqref{eq:superpotential},~\eqref{eq:soft_terms}
and~\eqref{eq:mass_basis}, we will leave understood (unless when
explicitly stated in the text) that all the other parameters
$Y_k,\,Z_k,\,B,\,m_2,\,$ etc. correspond to real and positive values.

Neglecting supersymmetry breaking effects in the RH sneutrino masses
and in the vertex, the total singlet sneutrino decay width is given by  
\begin{equation}
\Gamma_{\widetilde{N}_+}
=\Gamma_{\widetilde{N}_-}\equiv \Gamma_{\widetilde{N}}
=\frac {M}{4\pi} {\displaystyle \sum_k Y_{k}^2}\equiv
\frac {m_{\rm eff}\, M^2}{4 \pi\, v_u^2}, 
\label{eq:gamma}
\end{equation}
where $v_u$ is the vacuum expectation value of the up-type Higgs
doublet, $v_u=v\, \sin\beta $ (with $v$=174 GeV), and $m_{\rm eff}
\equiv \sum_k Y^2_k v_u^2/M$ is the $\tilde N_\pm$ decay parameter.
It is related with the washout parameter $K$ as
$K=\Gamma_{\widetilde{N}_+}/H(M)=m_{\rm eff}/m_*$ where the equilibrium
neutrino mass is $m_*=\sqrt{\frac{\pi g^*}{45}}\times\frac{8\pi^2
  v^2_u}{m_P}\sim 10^{-3}\,\,{\rm eV}$ with $g^*$ the total number
of relativistic degrees of freedom ($g^*=228.75$ in the MSSM)
and $m_P$ the Planck mass.

Equation (\ref{eq:soft_terms}) leads to three contributions to the CP
asymmetry in $\tilde N_{\pm} \rightarrow l_k \tilde h, \tilde l_k h $
decays \cite{soft3,ourgaugino}: $\epsilon_{k}^{S}$ arising from
self-energy diagrams induced by the bilinear $B$ term,
$\epsilon_{k}^{V}$ arising from vertex diagrams induced by the gaugino
masses, and $\epsilon_{k}^{I}$ which is due to interference of
self-energy and vertex.  They can be written as
\begin{eqnarray}
\epsilon_{k}^{S}\left(T\right) 
& = & - P_{k}  \, \frac{Z_k}{ Y_k }
\sin\phi_{Ak}\frac{A}{M}\frac{4B\Gamma}{4B^{2}+\Gamma^{2}}
\Delta_{BF}\left(T\right),  \label{eq:CP_asymres} \\
\epsilon_{k}^{V}\left(T\right)  
 & = & -P_k\,\frac{3\alpha_{2}}{4}\frac{m_{2}}{M}
\ln\frac{m_{2}^{2}}{m_{2}^{2}+M^{2}} \Delta_{BF}\left(T\right)\nonumber
\\
&& \times \left\{\frac{Z_k}{ Y_k } \left[ \sin\phi_{Ak}\frac{A}{M} 
\cos\left(2\phi_{g}\right)+\cos\phi_{Ak}\frac{A}{M}
\sin\left(2\phi_{g}\right)\right]
-\frac{B}{M}\sin\left(2\phi_{g}\right)\right\} 
, \label{eq:CP_asymver}  \\
\epsilon_{k}^{I}
\left(T\right) 
 & = &  P_{k} \frac{Z_k}{Y_k} \frac{3\alpha_{2}}{2}\sin\phi_{Ak} \frac{A}{M}
\left(\ln\frac{m_{2}^{2}}{m_{2}^{2}+M^{2}}\right)\cos\left(2\phi_{g}\right)
\frac{\Gamma^{2}}{4B^{2}+\Gamma^{2}}\Delta_{BF}\left(T\right),
\label{eq:CP_asymint} 
\end{eqnarray}
where
\begin{eqnarray}
\Delta_{BF}(T) & = & 
\frac{c^{s}(T) - c^{f}(T)}{c^{s}(T) + c^{f}(T)} 
\end{eqnarray}
is the thermal factor associated to the difference between the
phase-space factors for the scalar and fermionic channels, that
vanishes in the zero temperature limit $\Delta_{BF}(T\!=\!0)=0$.  As
long as we neglect the zero temperature slepton masses and
small Yukawa couplings, $c^f(T)$ and $c^s(T)$ are flavour independent
and they are the same for $\widetilde N_\pm$.  In the approximation in
which $\widetilde N_\pm$ decay at rest, the $c^{f,s}(T)$ functions are
given by: \bea c^f(T) &=&(1-x_{\ell}
-x_{\tilde{h}})\lambda(1,x_{\ell},x_{\tilde{h}}) \left[ 1-\fLeq\right]
\left[ 1-\fht\right],
\label{cfeq}\\
c^s(T)&=&\lambda(1,x_h,x_{\tilde{\ell}})
\left[ 1+\fh\right] \left[ 1+\fLteq\right],
\label{cbeq}
\eea
where
\bea
f^{eq}_{h,\tilde{\ell}}&=&\frac{1}{\exp[E_{h,\tilde{\ell}}/T]-1},
\label{eq:fHeq}\\
f^{eq}_{\tilde h,\ell}&=& \frac{1}{\exp[E_{\tilde h,\ell}/T]+1},  
\label{eq:fheq}
\eea are respectively the Bose-Einstein and Fermi-Dirac equilibrium
distributions, and \bea &E_{\ell,\tilde h}=\frac{M}{2}
(1+x_{\ell,\tilde{h}}- x_{\tilde h,\ell}), ~~~
E_{h,\tilde{\ell}}=\frac{M}{2} (1+x_{h ,\tilde{\ell}}-
x_{\tilde{\ell},h}),&\\
&\lambda(1,x,y)=\sqrt{(1+x-y)^2-4x},~~~ x_a\equiv
\frac{m_a(T)^2}{M^2}.& \eea The thermal masses for the relevant
supersymmetric degrees of freedom are \cite{thermal}: \bea m_h^2(T)=2
m_{\tilde h}^2(T)&=& \left(\frac{3}{8}g_2^2+\frac{1}{8}g_Y^2
  +\frac{3}{4}\lambda_t^2\right) \, T^2\; ,\\
m_{\tilde{\ell}}^2(T)=2 m_\ell^2(T)&=&
\left(\frac{3}{8}g_2^2+\frac{1}{8}g_Y^2 \right)\, T^2\; , \eea where
$g_2$ and $g_Y$ are the $SU(2)$ and $U(1)$ gauge couplings, and
$\lambda_t$ is the top Yukawa coupling, renormalized at the appropriate
energy scale.

In Eq.\eqref{eq:CP_asymres}  we have defined the Yukawa flavour projectors:
\begin{equation}
P_k \equiv  \frac{ Y_{k}^2}{\displaystyle 
\sum_j Y_{j}^2}
\end{equation}
which are  constrained by the condition
\begin{equation}
\sum_k P_k=1 \quad  \longrightarrow \quad 0\leq P_k\leq 1. 
\end{equation}

Regarding the flavour structure of the soft terms relevant 
for flavoured soft leptogenesis, we can distinguish two general 
possibilities:

1. {\it Universal soft supersymmetry breaking terms}. This case is
realized in supergravity and gauge mediated SUSY-breaking models (when
the renormalization group running of the parameters is neglected), and
in our notation corresponds to set
\begin{equation}
Z_k=  Y_k. 
\label{eq:uts}
\end{equation}
This {\sl Universal Trilinear Scenario} (UTS) is the one that was
considered so far in the literature on flavoured soft
leptogenesis~\cite{oursoft,ourgaugino}. In this case the only flavour
structure arises from the Yukawa couplings and both the total CP
asymmetries $\epsilon^k=\epsilon^S_k+\epsilon^V_k+\epsilon^I_k$ and
the corresponding washout terms, that are generically denoted as
$W_k$, are proportional to the same flavour projections, yielding:
\begin{equation}
\frac{\epsilon^e}{W_e}=
\frac{\epsilon^\mu}{W_\mu}=
\frac{\epsilon^\tau}{W_\tau}\; . 
\label{eq:epsWuts}
\end{equation}
Furthermore, as seen in Eq.\eqref{eq:CPphase1}
there is a unique phase for the trilinear couplings 
$\phi_{Ak}\equiv\phi_A=\arg(AB^*)$.

2. {\it General soft supersymmetry breaking terms}. In this case the
most general form for the soft-SUSY breaking terms is allowed, only
subject to the phenomenological constraints from limits on flavour 
changing neutral currents (FCNC) and from lepton flavour violating 
(LFV) processes. The trilinear soft-breaking terms are not aligned with
the corresponding Yukawa couplings, and Eq.~\eqref{eq:epsWuts} does
not hold. Studying this scenario can be rather involved due to the large
dimensionality of the relevant parameter space.  Therefore we will
introduce a drastic simplification that, while it can still capture
some of the main features of the general case, it allows to carry out
an analysis in terms of the same number of independent parameters than
in case 1.

Let us note that the CP asymmetries become flavour independent 
(except for the last term in Eq.\eqref{eq:CP_asymver})
if 
\begin{equation}
Z_{k}=\frac{\displaystyle
\sum_j |Y_{j}|^2}{3 Y^*_{k}}\; , 
\label{eq:ems}
\end{equation}
where we have kept $Z$ and $Y$ explicitly as complex numbers.
Eq.~\eqref{eq:ems} yields $\epsilon^k =\epsilon/3$ for each flavour,
and from Eq.~\eqref{eq:CPphase1} we see that, since $Z_kY^*_K$ is
real, also in this case there is a unique phase for the trilinear
couplings $\phi_{Ak}\equiv\phi_A=\arg(A B^*)$.  The normalization
factor of $1/3$ in Eq.~\eqref{eq:ems} has been introduced so that both
Eq.\eqref{eq:uts} and Eq.~\eqref{eq:ems} yield the same total
asymmetry $\sum_k\epsilon_k=\epsilon$. In what follows we will refer
to this case as the {\sl Simplified Misaligned Scenario} (SMS).  Our
SMS of course does not correspond to a completely general scenario,
and for example, due to the reduction in the number of independent
physical phases implied by Eq.~\eqref{eq:ems}, it excludes the
possibility of having flavour asymmetries of opposite signs, with
$|\epsilon^k| > |\epsilon|$ for some, or even for all, flavours. The
reader should thus keep in mind that enhancements of the final lepton
asymmetry even larger than the ones we will find within the SMS are
certainly possible.

\section{Lepton Flavour Equilibration}
\label{sec:lfe}

In the basis where charged lepton Yukawa couplings are diagonal, the SUSY
breaking slepton masses read
\begin{eqnarray}
\mathcal{L}_{soft} & \supset & \widetilde{m}_{ij}^{2}
\widetilde{\ell}_{i}^{*}\widetilde{\ell}_{j}.
\end{eqnarray}
The off-diagonal slepton masses $\widetilde{m}_{i\neq j}^{2}$ affect the
flavour composition of the slepton mass eigenstates so generically we
can write  
\begin{equation}
\widetilde{\ell}_{i}^{\left(int\right)}  = 
R_{ij}\widetilde{\ell}_{j} 
\end{equation}
where $R_{ij}$ is a unitary rotation matrix.  In this basis the
corresponding slepton-gaugino interactions in Eq.\eqref{eq:mass_basis} are
 \begin{eqnarray}
\nonumber 
\mathcal{L}_{\widetilde{\lambda},\tilde l} 
& =&
-g_{2}\left(\sigma_{1}\right)_{\alpha\beta}
\overline{\widetilde{\lambda}_{2}^{\pm}}P_{L}\ell_{i}^{\alpha}
R^*_{ij}\widetilde{\ell}_{j}^{\beta*}-\frac{g_{2}}{\sqrt{2}}
\left(\sigma_{3}\right)_{\alpha\beta}
\overline{\widetilde{\lambda}_{2}^{0}}
P_{L}\ell_{i}^{\alpha}R_{ij}^*\widetilde{\ell}_{j}^{\beta*}\nonumber
 \\  & &
-\frac{g_{Y}}{\sqrt{2}}\delta_{\alpha\beta}
\overline{\widetilde{\lambda}_{1}}
Y_{\ell}P_{L}  
\ell_{i}^{\alpha}R^*_{ij}\widetilde{\ell}_{j}^{\beta*} \,+\,\mbox{h.c.}\; ,
\label{eq:lgaugino}
 \end{eqnarray}
where $Y_\ell = -1$ is the hypercharge of the left-handed lepton doublets. 
The mixing matrix can be expressed in terms of the off-diagonal
slepton masses as:
\begin{eqnarray}
R_{ij} & \sim & \delta_{ij}+
\frac{\widetilde{m}_{ij}^{2}}{h_{i}^{2}T^{2}}
\nonumber \\
 & = & \delta_{ij}+\frac{\widetilde{m}_{ij}^{2}
v^2\cos^2\beta}{m_{i}^{2}M^{2}}z^{2},
\end{eqnarray}
where in the first line $h_{i}>h_{j}$ is the relevant charged Yukawa
coupling that determines at leading order the thermal mass splittings
of the sleptons, $v$ in the second line is the electroweak symmetry
breaking VEV with $v^2=v^2_u+v^2_d \simeq 174\,$GeV,
$z\equiv\frac{M}{T}$ where $T$ is the temperature and $M$ the mass of
the RH neutrino, and $m_i \equiv m_{\ell_i}(T\!=\!0)$ is the
zero temperature mass for the lepton $\ell_i$. 
In what follows, for simplicity we construct the $R_{ij}$ entries 
in such a way that they are flavour independent quantities.
We assume  $\widetilde{m}_{i\tau}=\widetilde{m}_{od}$ (for $i=e,\mu$) and 
$\widetilde{m}_{e\mu}=\widetilde{m}_{od} \frac{m_\mu}{m_\tau}$, where 
$\widetilde{m}_{od}$, is a unique off-diagonal soft-mass parameter. 
We thus obtain for $(ij)=(e\tau),\,(\mu\tau),\,(e\mu)$: 
\begin{eqnarray}
R_{ij} 
& \sim & \delta_{ij}+\frac{\widetilde{m}_{od}^{2}
\, v^2\cos^2\beta}{m_{\tau}^{2}M^{2}}z^{2},
\label{eq:rij}
\end{eqnarray}
where $m_{\tau}$ is the mass of the tau lepton.

$\mathcal{L}_{\widetilde{\lambda},\tilde l}$ in
Eq.~\eqref{eq:lgaugino} induces lepton flavour violating slepton
scatterings through the exchange of $SU(2)$ and $U(1)_Y$
gauginos. There are two possible t-channel scatterings
$\ell_{i}P\leftrightarrow\widetilde{\ell}_{j}\widetilde{P}$,
$\ell_{i}\widetilde{P}\leftrightarrow\widetilde{\ell}_{j}P$ and one
s-channel scattering $\ell_{i}\widetilde{\ell}_{j}^{*}\leftrightarrow
P\widetilde{P}^*$ (we denote $P$ as fermions and $\widetilde{P}$ as
scalars).  For processes mediated by $SU(2)$ gauginos
$P=\ell,q,\widetilde{h}$, while when mediated by $U(1)_Y$ gaugino one
must include the $SU(2)$ singlet states $P=e,u,d$ as well.
The corresponding reduced cross sections read: 
\begin{eqnarray}
\hat{\sigma}_{t1,G}^{ij}\left(s\right) & = & {\displaystyle \sum_P}
\frac{g_{G}^{4}\left|R_{ij}\right|^{2}\Pi_P^{G}}{8\pi}
\left[\left(\frac{2m_{\widetilde{\lambda}_{G}}^{2}}{s}+1\right)
\ln\left|\frac{m_{\widetilde{\lambda}_{G}}^{2}+s}
{m_{\widetilde{\lambda}_{G}}^{2}}\right|-2\right], \nonumber\\ 
\hat{\sigma}^{ij}_{t2,G}\left(s\right) 
 & = & {\displaystyle \sum_P}
\frac{g_{G}^{4}\left|R_{ij}\right|^{2}\Pi_P^{G}}{8\pi}
\left[\ln\left|\frac{m_{\widetilde{\lambda}_{G}}^{2}+s}
{m_{\widetilde{\lambda}_{G}}^{2}}\right|
-\frac{s}{m_{\widetilde{\lambda}_{G}}^{2}+s}\right], \nonumber\\ 
\hat{\sigma}^{ij}_{s,G}\left(s\right) 
 & = & {\displaystyle \sum_P}
\frac{g_{G}^{4}\left|R_{ij}\right|^{2}\Pi_P^G}
{16\pi}\left(\frac{s}{s-m_{\widetilde{\lambda}_{G}}^{2}}\right)^{2}  \; ,
\label{eq:sigLFE}
\end{eqnarray}
where $\Pi_{P}^{G}$ counts the numbers of degrees of freedom of the
particle P (isospin, quark flavours and color) involved in the scatterings
mediated by the $SU(2)$  ($G=2$) and $U(1)_Y$ ($G=Y$) gauginos respectively. 
In this last case the hypercharges $Y_\ell$ and $Y_P$ 
are also included in  $\Pi_{P}^{Y}$.
If the flavour changing scatterings in Eq.~\eqref{eq:sigLFE} are fast
enough, they will  lead to lepton flavour equilibration, and damp  
all  leptogenesis flavour effects \cite{lfe}
\footnote{See Ref.~\cite{juansheng} for some particular effects associated  
with lepton  flavour violating processes in scenarios with 
vanishing total CP asymmetry.}  .

The values of  $\widetilde{m}_{od}$ for which this occurs can be
estimated by comparing the  LFE scattering rates and
the $\Delta L=1$  washout rates.
Since the dominant $\Delta L=1$  contribution arises from inverse decays, 
the terms to be compared are: 
\begin{eqnarray}
\overline{\Gamma}_{\rm LFE}(T)
&\equiv&\frac{\gamma_{\rm LFE}(T)}{n^{c}_L(T)}\equiv
\frac{1}{n^{c}_L(T)} 
{\displaystyle \sum_{G,P}} \Pi^G_P
(\gamma^{ij}_{t1,G}+\gamma^{ij}_{t2,G}+\gamma^{ij}_{s,G})\nonumber \\
&= & \frac{1}{n^c_L(T)} \frac{T}{64 \pi}  
{\displaystyle \sum_{G}} 
\int ds \left[\hat \sigma^{ij}_{t1,G}(s)
+\hat \sigma^{ij}_{t2,G}(s)+\hat \sigma^{ij}_{s,G}(s)\right]\sqrt{s} 
{\cal K}_1\left(\frac{\sqrt{s}}{T}\right)\;, \\
\overline{\Gamma}_{\rm ID}(T)&\equiv& \frac{\gamma_{\tilde N}(T)}
{n^{c}_L(T)}= \frac{n_{\tilde N}^{eq}(T)}{n^{c}_L(T)}
\frac{{\cal K}_1(z)}{{\cal K}_2(z)} \Gamma_{\tilde N}\;,
\end{eqnarray}
where the $\gamma^{ij}$ in the first line represent the thermally 
averaged LFE reactions for one degree of freedom of the $P$-particle, 
$\mathcal{K}_{1,2}(z)$ are the modified Bessel function of 
the second kind of order 1 and 2,  
$\Gamma_{\widetilde{N}}$ is the zero temperature width   
Eq.~(\ref{eq:gamma}), and $n_{\tilde N}^{eq}$ is the equilibrium number
density for $\tilde N$  while $n^{c}_L=T^3/2$ is the relevant
density factor appearing in the washouts (see next section for
more details). In evaluating the
reaction densities above we have not included the thermal masses, and 
we have neglected Pauli-blocking
and stimulated emission as well as the relative motion of the particles
with respect to the plasma.

LFE scattering reaction densities have a different $T$ dependence with
respect to the Universe expansion and to the decay rates.  While for
the expansion $H(T) \sim T^2$, for LFE processes we have
$\overline{\Gamma}_{\rm LFE} \sim T^{-3}$.  This means that
the ratio $\overline{\Gamma}_{\rm LFE}/H\sim 1/T^5$, and thus once 
LFE reactions have attained thermal
equilibrium, they will remain in thermal equilibrium also at lower
temperatures.  In contrast, $\overline{\Gamma}_{\rm ID}$ first
increases till reaching a maximum, but then decreases exponentially
$\sim e^{-M/T}$ dropping out of equilibrium at temperatures not much
below $T\sim M$.  The relevant temperature where we should
compare the rates of these interactions is when the inverse decay rate
$\overline{\Gamma}_{\rm ID}$ becomes slower than the expansion rate of
the Universe $H$, that is when the lepton asymmetry starts being
generated from the out-of-equilibrium $\widetilde N_\pm$ decays.   We
define $z_{dec}$ as $\overline{\Gamma}_{\rm ID}(z_{dec})=H(z_{dec})$.
LFE is expected to be quite relevant for flavoured leptogenesis when
the following condition is verified:
\begin{eqnarray}
\overline{\Gamma}_{\rm LFE}\left(z_{dec}\right) &\geq& 
\overline{\Gamma}_{ID}\left(z_{dec}\right)
=H(z_{dec}).
\end{eqnarray}
In this case, LFE processes are in equilibrium since the very onset of
the era of out-of-equilibrium decays, and due to its temperature
dependence it is guaranteed that they will remain in equilibrium until
leptogenesis is over.

\begin{figure}[htb]
\epsfig{file=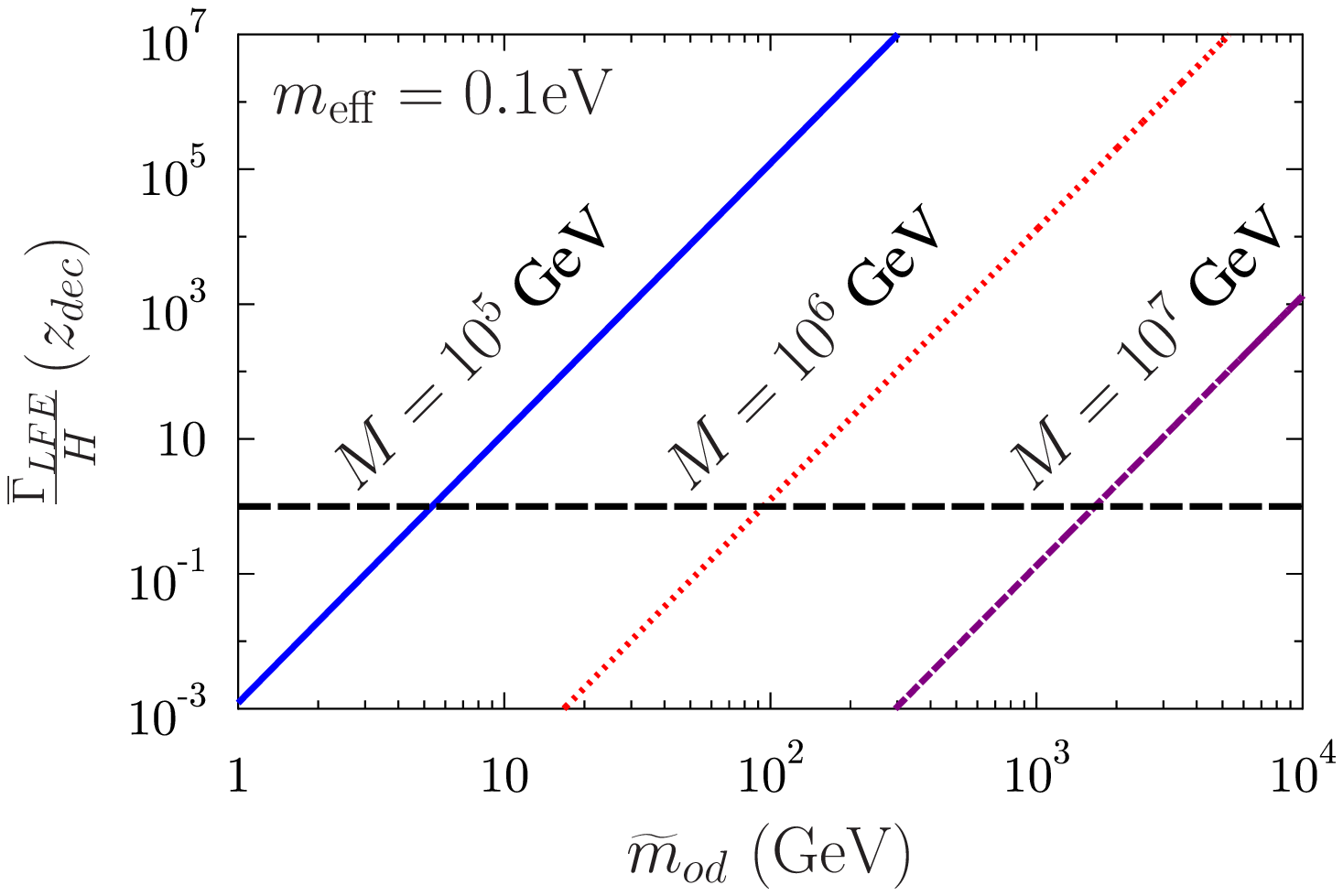,width=0.5\textwidth,angle=0}
\hspace*{-.2cm} 
\epsfig{file=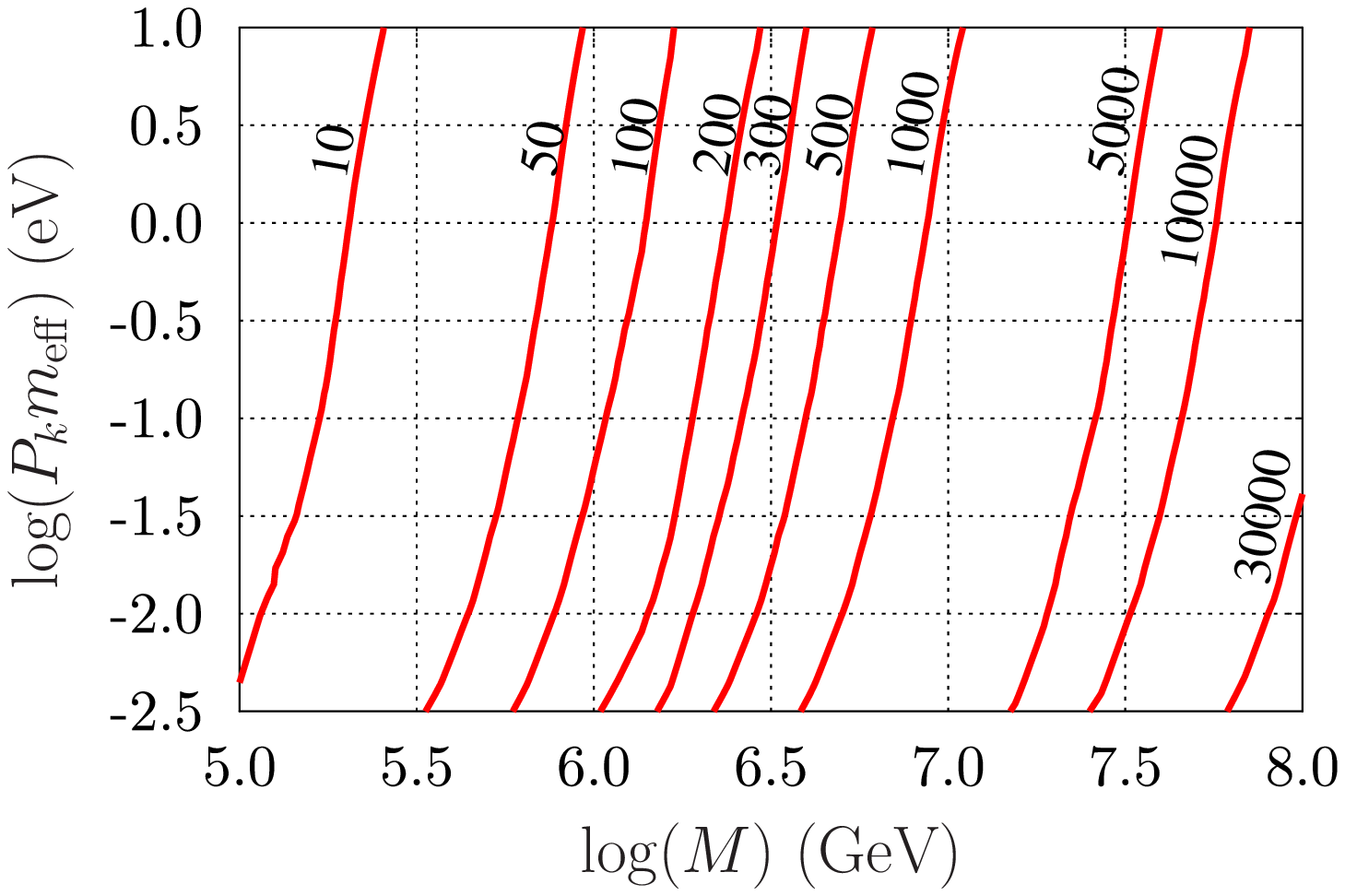,width=0.5\textwidth,angle=0}
\caption{ The left panel shows the ratio of $\bar \Gamma_{\rm LFE}$ to
  the Hubble expansion rate $H$ at $z_{dec}$ as a function of
  $\widetilde{m}_{od}$ for $m_{\rm eff}=0.1$ eV and $\tan\beta=30$ and
  three values of $M$.  The right panel shows in the ($P_k m_{\rm
    eff}, M$) plane, contours of constant values of
  $\widetilde{m}_{od}$ (in GeV) for which
  $\overline{\Gamma}_{\rm LFE} \left(z_{dec}^k\right) \geq
  P_k\overline{\Gamma}_{\rm ID}\left(z_{dec}^k\right)$.}
\label{fig:lfe}
\end{figure}

In the left panel of Fig.~\ref{fig:lfe} we plot the ratio
$\overline{\Gamma}_{\rm ID}(z_{dec})/H(z_{dec})$ as a function of
$\widetilde{m}_{od}$ for $m_{\rm eff}=0.1$ eV (defined in
Eq.~\eqref{eq:gamma}), $\tan\beta=30$, and for different values of $M$.
From the figure we can read the characteristic value of
$\widetilde{m}_{od}$ for which LFE becomes relevant.  Notice that the
dominant dependence on $\tan\beta\sim 1/\cos\beta $ ($\tan\beta \gg
1$) arises due to $v\cos\beta=v_d$ in Eq.~\eqref{eq:rij}. Thus the
results from other values of $\tan\beta$ can be easily read from the
figure by rescaling 
$\widetilde{m}^\beta_{od}=\widetilde{m}^{\rm fig}_{od}/(30 \cos\beta)$.

Since we are interested in the dynamics of lepton flavours, to be more
precise about LFE effects we should in fact consider the temperature
$z_{dec}^k$ at which the inverse decay rate for one specific flavour
goes out of equilibrium, that can be defined through
$\overline{\Gamma}_{\rm ID}^k(z_{dec}^k)=P_k\overline{\Gamma}_{\rm
  ID}(z_{dec}^k) = H(z_{dec}^k)$. Let's assume $P_{a}<P_{b}<P_{c}$ which
implies $z_{dec}^{a}<z_{dec}^{b}<z_{dec}^{c}$.  In other words,
assuming that the lepton doublet $\ell_{a}$ is the most weakly coupled
to $\widetilde N_\pm$, $\overline{\Gamma}_{\rm ID}^{a}$ will go out of
equilibrium first, and then $\overline\Gamma_{\rm ID}^{b}$ and
$\overline{\Gamma}_{\rm ID}^{c}$ will follow.  Hence, for given values
of $m_{\rm eff}$ and $M$, the minimum value $\widetilde{m}_{od}^{min}$
for which LFE effects start being important is given by the following
condition:
\begin{eqnarray}
\overline{\Gamma}_{\rm LFE}
\left(z_{dec}^{c}\right) &\simeq& 
\overline{\Gamma}_{\rm ID}^{c}\left(z_{dec}^c
\right)
\;\;\;
\Rightarrow 
\mbox{\ \ determines\ \  $\widetilde{m}_{od}^{min}$}, 
\label{eq:LFE_condition1} \; .
\end{eqnarray}
For $\widetilde{m}_{od} \ll \widetilde{m}_{od}^{min}$ 
LFE effects can
be neglected, since they will attain thermal equilibrium only after
leptogenesis is completed.

In the right panel in  Fig.~\ref{fig:lfe}, we plot 
in  the plane of the flavoured effective decay mass 
$P_k m_{\rm eff}$ and of the RH sneutrino mass $M$,
various contours corresponding to different values 
of $\widetilde{m}_{od}$  for which 
$\overline{\Gamma}_{\rm LFE}\left(z_{dec}^k\right)=
P_k\Gamma_{\rm ID}\left(z_{dec}^k\right)$.
For a given value of $M$ and $m_{\rm eff}$, 
and  for a given set of flavour projections $P_a<P_b< P_c$, 
$\widetilde{m}_{od}^{min}$ 
is given by the value of the 
$\widetilde{m}_{od}$  curve for  which the vertical line  $x=M$ 
intersects  the corresponding contour at  $y_c=P_c\, m_{\rm eff}$. 

Furthermore, since $\overline{\Gamma}_{\rm LFE}$ has a rather strong
dependence on $\widetilde{m}_{od}$ ($ \overline{\Gamma}_{\rm
  LFE}\propto \widetilde{m}_{od}^4$), one expects that the value
$\widetilde{m}_{od}^{max}$ for which LFE effects completely
equilibrate the asymmetries in the different lepton flavours will not
be much larger than $\widetilde{m}_{od}^{min}$. Indeed our numerical
results (see next section) show that $\widetilde{m}_{od}^{max} \sim
{\cal O}(5-10)\, \widetilde{m}_{od}^{min}$. Clearly, 
as far as
leptogenesis is concerned,  
larger values
$\widetilde{m}_{od}\gg \widetilde{m}_{od}^{max} \sim
\widetilde{m}_{od}^{min}$ do not imply any modification in the numerical 
results with respect to what is obtained with 
$\widetilde{m}_{od} = \widetilde{m}_{od}^{max}$.

\section{Results}
\label{sec:results}
We quantify the results that can be obtained by including LFE effects
by solving the following set of BE for the abundances $Y_X=n_X/s$:
\begin{eqnarray}
\label{eq:N}
-sHz\frac{dY_{N}}{dz} & \!= \! & \left(\frac{Y_{N}}{Y_{N}^{eq}}
-1\right)\left(\gamma_{N}+4\gamma_{t}^{(0)}+4\gamma_{t}^{(1)}
+4\gamma_{t}^{(2)}+2\gamma_{t}^{(3)}+4\gamma_{t}^{(4)}\right),\qquad\qquad
\mbox{\hspace{1.2cm} \phantom{i}}
\\
\label{eq:tildeN}
-sHz\frac{dY_{\widetilde{N}_{tot}}}{dz} & \! =\! & 
\left(\frac{\gamma_{\widetilde{N}}}{2}+
\gamma_{\widetilde{N}}^{\left(3\right)}+
3\gamma_{22}+2\gamma_{t}^{\left(5\right)}+
2\gamma_{t}^{\left(6\right)}+2\gamma_{t}^{\left(7\right)}+
\gamma_{t}^{\left(8\right)}+2\gamma_{t}^{\left(9\right)}\right)
\left(\frac{Y_{\widetilde{N}_{tot}}}{Y_{\widetilde{N}}^{eq}}-2\right) 
\end{eqnarray}
\begin{eqnarray}
-sHz\frac{dY_{\Delta_{tot}^{k}}}{dz} & = & 
\epsilon^{k}\left(T\right)\frac{\gamma_{\widetilde{N}}}{2}
\left(\frac{Y_{\widetilde{N}_{tot}}}{Y_{\widetilde{N}}^{eq}}-2\right)
-\left[\frac{\gamma^k_{\widetilde{N}}}{2}+\frac{\gamma^k_{N}}{2}+
\gamma_{\widetilde{N}}^{\left(3\right)k}+
\left(\frac{1}{2}\frac{Y_{\widetilde{N}_{tot}}}{Y_{\widetilde{N}}^{eq}}
+2\right)\gamma^k_{22}\right]
\left(\frac{Y_{L^k_{tot}}}{Y_{L_{tot}}^{c}}+
\frac{Y_{H_{tot}}}{Y_{H_{tot}}^{c}}\right)
\nonumber \\
 &  & -2\left(\gamma_{t}^{\left(1\right)k}+\gamma_{t}^{\left(2\right)
k}+\gamma_{t}^{\left(4\right)k}+\gamma_{t}^{\left(6\right)k}
+\gamma_{t}^{\left(7\right)k}+\gamma_{t}^{\left(9\right)k}\right)
\frac{Y_{L_{tot}^{k}}}{Y_{L_{tot}}^{c}}
\nonumber \\
 &  & -\left[\left(2\gamma_{t}^{\left(0\right)}+\gamma_{t}^{\left(3\right)
k}\right)\frac{Y_{N}}{Y_{N}^{eq}}+\left(\gamma_{t}^{\left(5\right)k}+
\frac{1}{2}\gamma_{t}^{\left(8\right)k}\right)
\frac{Y_{\widetilde{N}_{tot}}}{Y_{\widetilde{N}}^{eq}}\right]
\frac{Y_{L_{tot}^{k}}}{Y_{L_{tot}}^{c}}
\nonumber \\
 &  & -\left(2\gamma_{t}^{\left(0\right)k}+\gamma_{t}^{\left(1\right)
k}+\gamma_{t}^{\left(3\right)k}+\gamma_{t}^{\left(4\right)k}
+2\gamma_{t}^{\left(5\right)k}+\gamma_{t}^{\left(6\right)k}
+\gamma_{t}^{\left(7\right)k}+\gamma_{t}^{\left(8\right)k}
+\gamma_{t}^{\left(9\right)k}\right)\frac{Y_{H_{tot}}}{Y_{H_{tot}}^{c}}
\nonumber \\
 &  & -\left[\left(\gamma_{t}^{\left(1\right)k}+
\gamma_{t}^{\left(2\right)k}+\gamma_{t}^{\left(4\right)k}\right)
\frac{Y_{N}}{Y_{N}^{eq}}+\frac{1}{2}\left(\gamma_{t}^{\left(6\right)k}+
\gamma_{t}^{\left(7\right)k}+\gamma_{t}^{\left(9\right)k}\right)
\frac{Y_{\widetilde{N}_{tot}}}{Y_{\widetilde{N}}^{eq}}\right]
\frac{Y_{H_{tot}}}{Y_{H_{tot}}^{c}}
\nonumber \\
&&
-84 \sum_{j\neq k}
\left(\gamma^{jk}_{t1,2}+\gamma^{jk}_{t2,2}+\gamma^{jk}_{s,2}\right)
\frac{Y_{L^k_{tot}}-Y_{L^j_{tot}}}{Y^c_{L_{tot}}}\nonumber \\
 &  & -72 \sum_{j\neq k}\left(\gamma^{jk}_{t1,Y}+\gamma^{jk}_{t2,Y}
+\gamma^{jk}_{s,Y}\right)
\frac{Y_{L^k_{tot}}-Y_{L^j_{tot}}}{Y^c_{L_{tot}}}. 
\label{eq:BE_fla_tot}
\end{eqnarray}
Spectator effects and sphaleron flavour mixing are taken into account by writing
\begin{equation}
\label{eq:leptonHiggs}
Y_{L_{tot}^{k}} =  \sum_{j}A_{kj}
Y_{\Delta_{tot}^{j}},\qquad\qquad 
Y_{H_{tot}} = \sum_{j}C_{j}Y_{\Delta_{tot}^{j}}.
\end{equation}
In Eqs.~\eqref{eq:N}-\eqref{eq:BE_fla_tot} we have defined
$Y_{\Delta^k_{tot}} \equiv Y_B/3 - Y_{L^k_{tot}}$,
$Y_{\widetilde{N}_{
    {tot}}}\equiv Y_{\widetilde{N}_{+}}+Y_{\widetilde{N}_{-}}$, and
$Y_{L^k_{
    {tot}}} \equiv Y_{L^k_{f}}+Y_{L^k_{s}}$ that is the total
asymmetry in flavour $k$ obtained by summing up both fermions
$Y_{L^k_f}$ and scalars $Y_{L^k_s}$ contributions, where for example
$Y_{L^k_f}=(Y_{\ell_k}-Y_{\bar\ell_k})$, while $Y_{H_{{tot}}}$ is
the total asymmetry for the Higgs and Higgsinos.  In addition we have
$Y_{H_{tot}}^{c} =Y_{L_{tot}}^{c}\equiv\frac{45}{4\pi^{2}g^*}$ and
$Y^{\rm eq}_{\tilde N}(T\gg M) = 90 \zeta(3)/(4\pi^4g^*)$.

The values of the entries of the matrix $A$ and of the vector $C$ in
Eq.~\eqref{eq:leptonHiggs} depend on the range of temperature, that is
on the particular set of interactions that are in equilibrium when
leptogenesis is taking place. For
$T<(1+\tan^{2}\beta)\times10^{5}\mbox{GeV}$ reactions mediated by the
Yukawa couplings of all the three families are in equilibrium\cite{leptoreview}, and in
this case the $A$ and $C$ matrices are given by~\footnote{Indeed we
  find that, within a given $T$ regime, $A$ and $C$ for the MSSM and
  for the SM are the same up to a global factor $1/2$ for $C$.  This
  is expected to be so, since supersymmetry cannot alter the flavour
  distribution between the charges.  This is in agreement with the
  analysis in Ref.\cite{inui}, but it disagrees with the $A$ matrix
  given in Ref.\cite{antusch}.}
\begin{eqnarray}
A=\frac{2}{711}\left(\begin{array}{ccc}
-221 & 16 & 16\\
16 & -221 & 16\\
16 & 16 & -221
\end{array}\right), &\;\;\;\;\;& C=-\frac{8}{79}\left(1\;\;1\;\;1\right).
\label{eq:AC_Matrix}
\end{eqnarray}
Note that the last two lines in Eq.\eqref{eq:BE_fla_tot} correspond to the
reaction densities for the LFE processes given in 
Eq.\eqref{eq:sigLFE}, and play the role of 
controlling the effectiveness of the leptogenesis  
flavour effects.  The different reaction densities 
 for the  $\Delta L=1$  processes are:
\begin{eqnarray}
&&\gamma_{\widetilde N}^k= 
{\displaystyle \sum_{i=\pm}}
\left[
\gamma(\widetilde{N}_{i}\leftrightarrow
\bar{\tilde{h}}\ell_k)
+\gamma(\widetilde{N}_{i} \leftrightarrow h\tilde{\ell_k})\right] ,
\nonumber \\
&&\gamma^{(3)k}_{\widetilde N}=
{\displaystyle \sum_{i=\pm}}
\gamma( 
\widetilde{N}_{i}\leftrightarrow 
\tilde{\ell_k}^{*}\tilde{u}\tilde{q})\; , 
\nonumber\\
&&\gamma_{22}^k =  {\displaystyle \sum_{i=\pm}}
\gamma(\widetilde{N}_{i}
\tilde{\ell_k}\leftrightarrow\tilde{u}\tilde{q}
)=
{\displaystyle \sum_{i=\pm}}
\gamma(\widetilde{N}_{i}
\tilde{q}^{*}\leftrightarrow\tilde{\ell_k}^{*}\tilde{u}
)=
{\displaystyle \sum_{i=\pm}}
\gamma(\widetilde{N}_{i}\tilde{u}^{*}\leftrightarrow\tilde{\ell_k}^{*}\tilde{q}) ,
\nonumber \\
&&\gamma_t^{(5)k}={\displaystyle \sum_{i=\pm}}
\gamma(\widetilde{N}_{i}
\ell_k\leftrightarrow q\tilde{u})={\displaystyle \sum_{i=\pm}}\gamma(
\widetilde{N}_{i}\ell_k\leftrightarrow\tilde{q}\bar{u})\; ,
\nonumber \\
&&\gamma_t^{(6)k}={\displaystyle \sum_{i=\pm}}
\gamma(
\widetilde{N}_{i}\tilde{u}^{*}\leftrightarrow\bar{\ell_k}q)=
{\displaystyle \sum_{i=\pm}}
\gamma( 
\widetilde{N}_{i}\tilde{q}^{*}\leftrightarrow\bar{\ell_k}\bar{u})\;,
\nonumber \\
&&\gamma_t^{(7)k}= {\displaystyle \sum_{i=\pm}}
\gamma(
\widetilde{N}_{i}\bar{q}\leftrightarrow\bar{\ell_k}\tilde{u})={\displaystyle \sum_{i=\pm}}\gamma( 
\widetilde{N}_{i}u\leftrightarrow\bar{\ell_k}\tilde{q}) , 
\nonumber \\
&&\gamma_t^{(8)k}= {\displaystyle \sum_{i=\pm}}
\gamma(
\widetilde{N}_{i}\tilde{\ell_k}^{*}\leftrightarrow\bar{q}u) ,
\nonumber \\
&&\gamma_t^{(9)k}={\displaystyle \sum_{i=\pm}}
\gamma(
\widetilde{N}_{i}q\leftrightarrow\tilde{\ell_k}u)= 
{\displaystyle \sum_{i=\pm}}
\gamma(\widetilde{N}_{i}\bar{u}\leftrightarrow\tilde{\ell_k}\bar{q}) ,
\nonumber \\
&&\gamma_N^k=\gamma(N\leftrightarrow \ell_k h)+
\gamma(N\leftrightarrow \tilde{\ell_k}^* \tilde h) ,
\nonumber \\
&&\gamma_t^{(0)k}=\gamma(N\tilde{\ell_k}\leftrightarrow q\tilde{u})=
\gamma(N\tilde{\ell_k}\leftrightarrow\tilde{q}\bar{u}) ,
\nonumber \\
&&\gamma_t^{(1)k}=\gamma(N\bar{q}\leftrightarrow\tilde{\ell_k}^{*}\tilde{u})=
\gamma(Nu\leftrightarrow\tilde{\ell_k}^{*}\tilde{q})\; ,
\nonumber \\
&&\gamma_t^{(2)k}=\gamma(N\tilde{u}^{*}\leftrightarrow\tilde{\ell_k}^{*}q)=
\gamma(N\tilde{q}^{*}\leftrightarrow\tilde{\ell_k}^{*}\bar{u})\; , 
\nonumber\\
&&\gamma_t^{(3)k}=\gamma(N\ell_k\leftrightarrow q\bar{u})\; ,
\nonumber \\
&&\gamma_t^{(4)k}=\gamma (Nu\leftrightarrow\bar{\ell_k}q)=
\gamma(N\bar{q}\leftrightarrow\bar{\ell_k}\bar{u})\; .
\label{eq:gammas}
\end{eqnarray}
When no flavour index appears in the $\gamma$'s, 
as in Eqs.~\eqref{eq:N} and \eqref{eq:tildeN}, 
it is understood that the corresponding reactions have 
been  summed over all lepton flavours.
The flavoured and flavour-summed rates are thus related 
according to 
\begin{equation}
\label{eq:flavouredrate}
\gamma_X^{k} =P_k \gamma_X\; .
\end{equation}
In our calculation we have kept particle thermal masses and
fermion-boson statistical factors only in the CP asymmetries, but we
have neglected them in the rest of the reaction rates, with the
exception of the thermal Higgs mass in the $\Delta L=1$ processes
involving a Higgs boson exchange in the $t$-channel, that is needed to
regulate the IR divergence that occurs in the limit $m_h \to 0$.

In what fallows we consider the resonant (self-energy) CP asymmetry
$\epsilon_{k}^{S}\left(T\right)$.
We parametrize the asymmetry generated by the decay of the 
singlet sneutrino states in a given flavour as 
\begin{equation}
Y_{\Delta^k_{tot}}(z\rightarrow \infty)=-2 \eta_k\, \bar\epsilon  \, 
Y^{\rm eq}_{\widetilde N}(T\gg M) 
\label{eq:yb-lfla}
\end{equation}
where, to facilitate  comparison with the existing literature,
we have defined  
\begin{equation}
\bar\epsilon=-\sin\phi_A
\frac{A}{M}\frac{4B\Gamma}{4B^{2}+\Gamma^{2}}\; .
\label{eq:CP_asym}
\end{equation}

The final amount of ${B}-{L}$ asymmetry generated in the decay of the 
singlet sneutrinos (we assume no pre-existing
asymmetry) can be parametrized as:
\begin{equation}
Y_{B-L}(z\rightarrow \infty)=\sum_k Y_{\Delta^k_{tot}}(z\rightarrow \infty)=
-2 \eta\, \bar\epsilon \, Y^{\rm eq}_{\widetilde N}(T\gg  M), 
\label{eq:yb-l}
\end{equation}
with 
\begin{equation}
\eta=\sum_k \eta_k .
\end{equation}

After conversion by the sphaleron transitions, the final baryon asymmetry
is related to the ${B}-{L}$ asymmetry by
\begin{equation}
Y_{B}= \frac{24+4n_h}{66+13n_h}
Y_{B-L}(z\rightarrow \infty)
=\frac{8}{23} \, Y_{B-L}(z\rightarrow \infty) \; , 
\label{eq:yb}
\end{equation}
where $n_h$ is the number of Higgs doublets, and in the second
equality we have taken $n_h=2$.

\begin{figure}[htb]
\epsfig{file=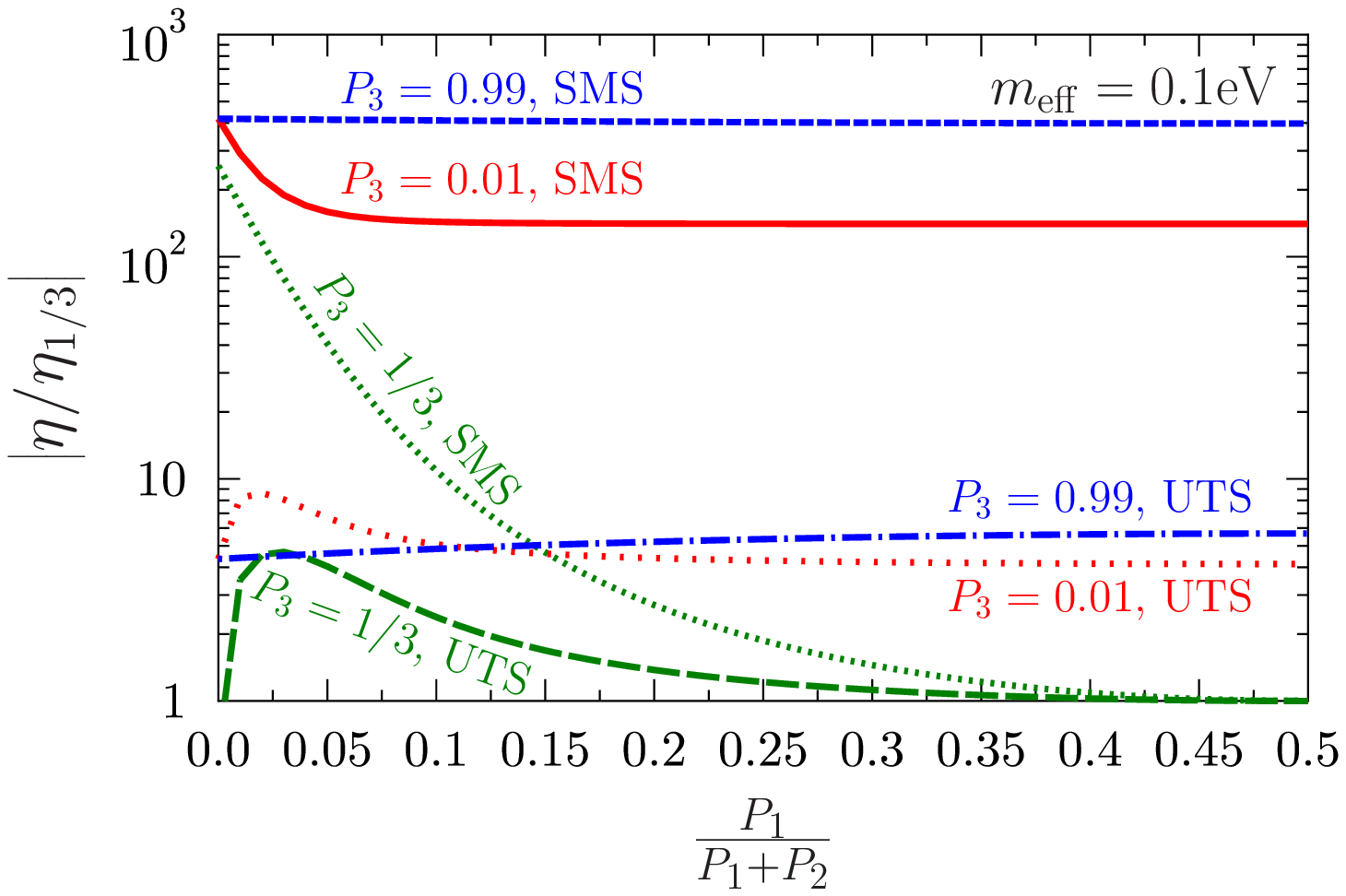,width=0.5\textwidth,height=0.25\textheight,angle=0} 
\epsfig{file=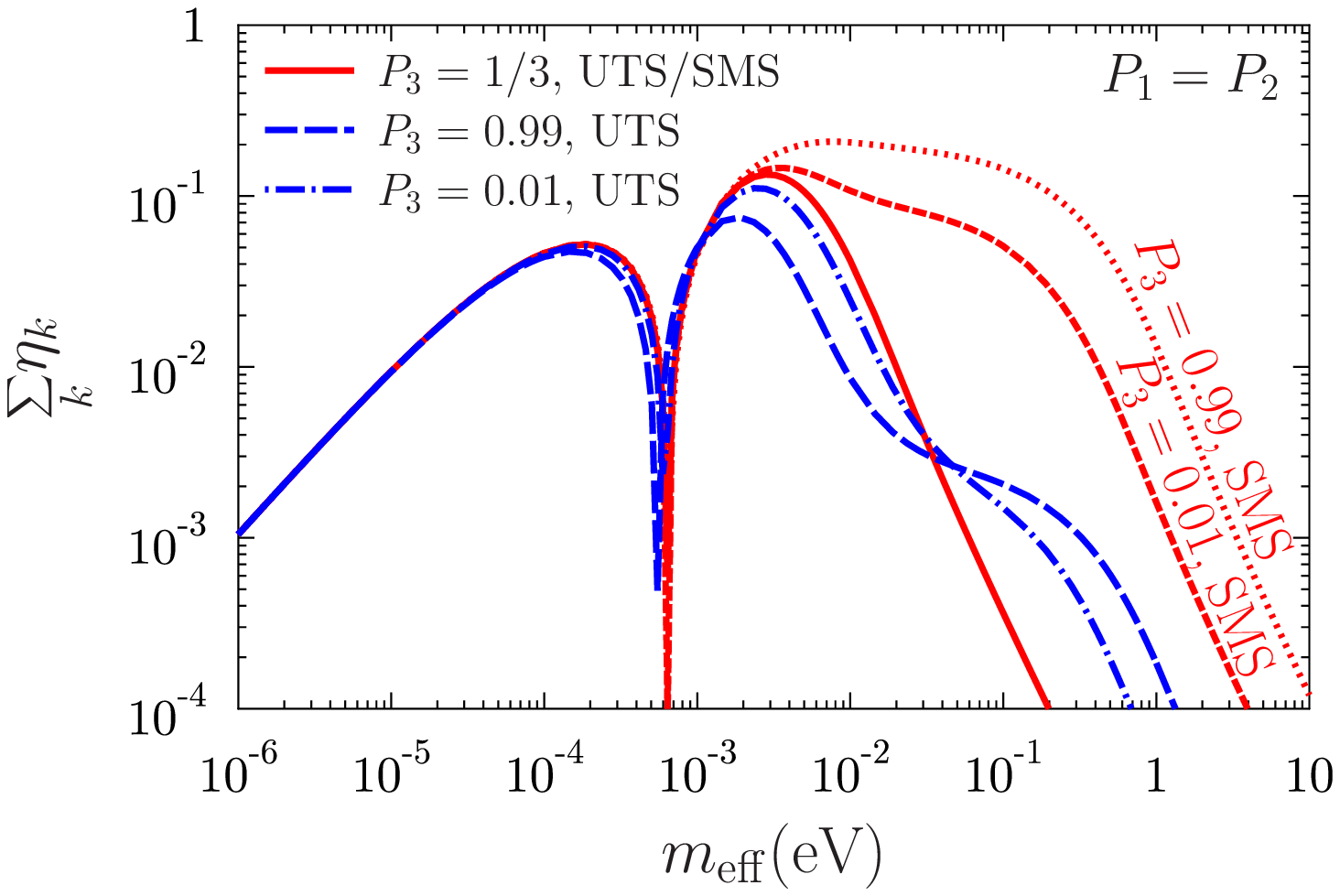,width=0.5\textwidth,,height=0.25\textheight,angle=0} 
\caption{
The dependence of the efficiency -- normalized to the flavour
equipartition case $P=(1/3,1/3,1/3)$ -- on the values of the lepton flavour 
projections (left) and on $m_{\rm eff}$ (right). The figures correspond to   
$M=10^6$ GeV and $\tan\beta=30$.}
\label{fig:eff_fla}
\end{figure}

According to our expressions for the CP asymmetries
Eqs.~\eqref{eq:CP_asymres}-\eqref{eq:CP_asymint} and to the general
expression for the flavoured reaction rates
Eq.~\eqref{eq:flavouredrate}, and neglecting for the time being LFE
effects, $\eta_k$ depends on flavour via the projections $P_k$ and
$Z_k/Y_k$. The final asymmetry produced also depends on the Yukawa 
couplings $\sum_j Y_j^2$ and on the heavy singlet mass $M$
through the combination $m_{\rm eff}$ defined in 
 Eq.~\eqref{eq:gamma}
(there is a residual mild dependence on $M$ due to the running of 
the top Yukawa coupling).
The dependence of the efficiency factor on the flavour projections and
on $m_{\rm eff}$ is shown in Fig.\ref{fig:eff_fla}.  The results are
obtained assuming that the ${\tilde N}$ population is created by their
Yukawa interactions with the thermal plasma, that is $Y_{\tilde
  N}(z\rightarrow 0)=0$, and neglecting for the time being the
possible effects of LFE. The plot is shown for $M=10^6$ GeV and
$\tan\beta=30$ although, as mentioned above, the efficiency is
practically independent of $M$.  As long as $\tan\beta$ is not very
close to one, the dominant dependence on $\tan\beta$ arises via $v_u$
as given in Eq.~(\ref{eq:gamma}) and it is therefore also rather
mild. For $\tan\beta\sim {\cal O}(1)$ there is also an additional
(very weak) dependence due to the associated change in the top Yukawa
coupling.  The results are displayed for the two choices of
soft-breaking terms that have been discussed at the end of
Sec.~\ref{sec:lag}: the UTS, defined by Eq.\eqref{eq:uts}, and our
SMS, defined by Eq.\eqref{eq:ems}.  We note that these two scenarios
are equivalent for the special case of flavour equipartition 
$P_1=P_2=P_3=1/3$.

From the left panel in Fig.\ref{fig:eff_fla} we see that departure
from the equipartition flavour case results in an enhancement of the
efficiency, and that particularly large enhancements are possible for
the SMS scenario.  Note that the top line in the left panel of
  Fig.~\ref{fig:eff_fla} labeled $P_3=0.99$ represents the maximum
  enhancement that can be obtained in the SMS (relaxing the constraint
  in Eq.~\eqref{eq:ems} that defines our SMS, larger enhancements are
  however possible).  This is because for $P_3=0.99$ both the
  asymmetries $Y_{\Delta^1_{tot}}$ and $Y_{\Delta^2_{tot}}$ are
  generated in the weak washout regime, that is, approximately within
  the same temperature range, and in the SMS this implies $\epsilon_1
  (T_1) \approx \epsilon_2 (T_2)$. The related combined efficiency is
  then simply determined by $(P_1+P_2)\,m_{\rm eff}\simeq m_*$ and is
  thus always maximal, independently of the individual values of $P_1$
  and $P_2$, as is apparent from the figure.

The right panel  of Fig.~\ref{fig:eff_fla} 
shows the dependence of the efficiency  on $m_{\rm eff}$ in the flavour
equipartition case and for two other sets of flavour projections. As
it is known, flavour effects become more relevant when the washouts get
stronger.  This is confirmed in this picture where it is seen that for
the SMS scenario the possible enhancements quickly grow with $m_{\rm
  eff}$.  Note that in soft leptogenesis this dependence is even
stronger than in standard leptogenesis. This is due to the fact that
the flavoured washout parameters $P_k m_{\rm eff}$ also determine the
value of $z^k_{dec}$ when the lepton asymmetry in the $k$ flavour
starts being generated, and since the CP asymmetry has a strong
dependence on $z$, different values of $P_1,\,P_2,\,$ and $P_3$ imply
that the corresponding flavour asymmetries are generated with
different values of the CP asymmetry even when, as in the SMS, the
fundamental quantity $\bar \epsilon$ is flavour independent.  In
summary, what happens is that the flavour that suffers the weakest
washout is also the one for which inverse-decays go out of equilibrium
earlier, and thus also the one for which the lepton asymmetry starts
being generated when $\bar\epsilon\times\Delta_{BF}$ is larger. This
realizes a very efficient scheme in which the flavour that is more
weakly washed out has effectively the largest CP asymmetry, and this
explains qualitatively the origin of the large enhancements that we
have found. Furthermore,  when $P_k m_{\rm eff}\ll m_*$ 
so that the inverse decay of flavour $k$  never reaches  equilibrium 
and the washout of the asymmetry $Y_{\Delta^k_{tot}}$ is negligible,
the maximum efficiency is reached.

We should however spend a word of caution for the reader about
interpreting our numerical results in the weak washout regime and, for
the SMS, also in the limit of extreme flavour hierarchies ($P_k \to
0$).  At high temperatures $(z < 1)$ the Higgs bosons (higgsinos)
develop a sufficiently large thermal mass to decay into sleptons
(leptons) and sneutrinos.  The new CP asymmetries associated with these decays
could be particularly large~\cite{thermal}, and thus sizable lepton
flavour asymmetries could be generated at high temperatures. 
This type of thermal effects are not included in our analysis.  
Concerning the flavour decoupling limit
within the SMS, clearly when $P_k \to 0$ no asymmetry can be generated
in the flavour $k$.  However, in our SMS flavour asymmetries are
defined to be independent of the projectors $P$ and thus survive in
the $P\to 0$ limit. On physical grounds, one would expect for example
that when one decay branching ratio is suppressed, say, as $P <
10^{-5}$, the associated CP asymmetry will be at most of $ {\cal
  O}(10^{-7})$ and thus irrelevant for leptogenesis.
This means that for extreme flavour hierarchies, the SMS breaks down
as a possible physical realization of soft leptogenesis, and thus in
what follows we will restrict our considerations to a range of
hierarchies $P\gsim 10^{-3}$.

As a result of our analysis, we find that for the SMS scenario with
hierarchical Yukawa couplings, successful leptogenesis is possible
even for $m_{\rm eff}\gg {\cal O} ({\rm eV})$.  For example, as is
shown in the right
panel of Fig.~\ref{fig:eff_fla}, for $P_1=P_2=5 \times 10^{-3}$ and
$m_{\rm eff}\sim 5$ eV, we obtain $|\eta|\sim  10^{-3}$, that
yields the estimate
\begin{equation}
Y_{B} ({\rm SMS}, P_1=P_2=5 \times 10^{-3},m_{\rm eff}=5\;{\rm  eV})\sim 10^{-6} \times 
\overline\epsilon \;. 
\end{equation}
Thus we see that assuming a large, but still acceptable value of $\bar\epsilon
\sim 10^{-4}$, soft leptogenesis can successfully generate the observed
baryon asymmetry~\cite{lastwmap}:
\begin{equation}
Y_{B_{obs}}= (8.78\pm0.24)\times 10^{-11}\;  
\label{eq:wmapeta}
\end{equation}
also for values of  $m_{\rm eff}$ that are about two orders of 
magnitude larger 
than what is found in the unflavoured standard  leptogenesis scenario.

\begin{figure}[htb]
\epsfig{file=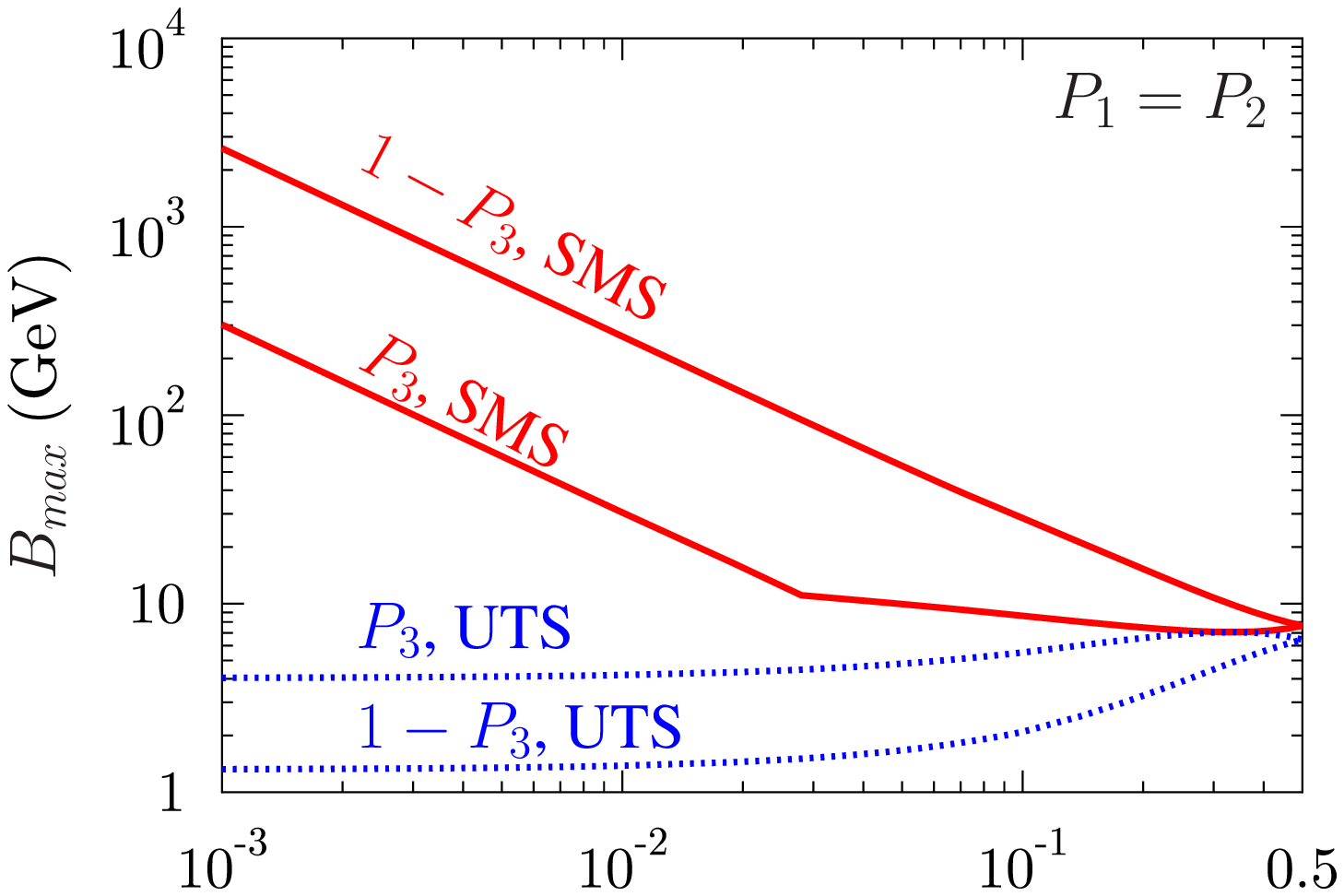,width=0.5\textwidth,angle=0}
\epsfig{file=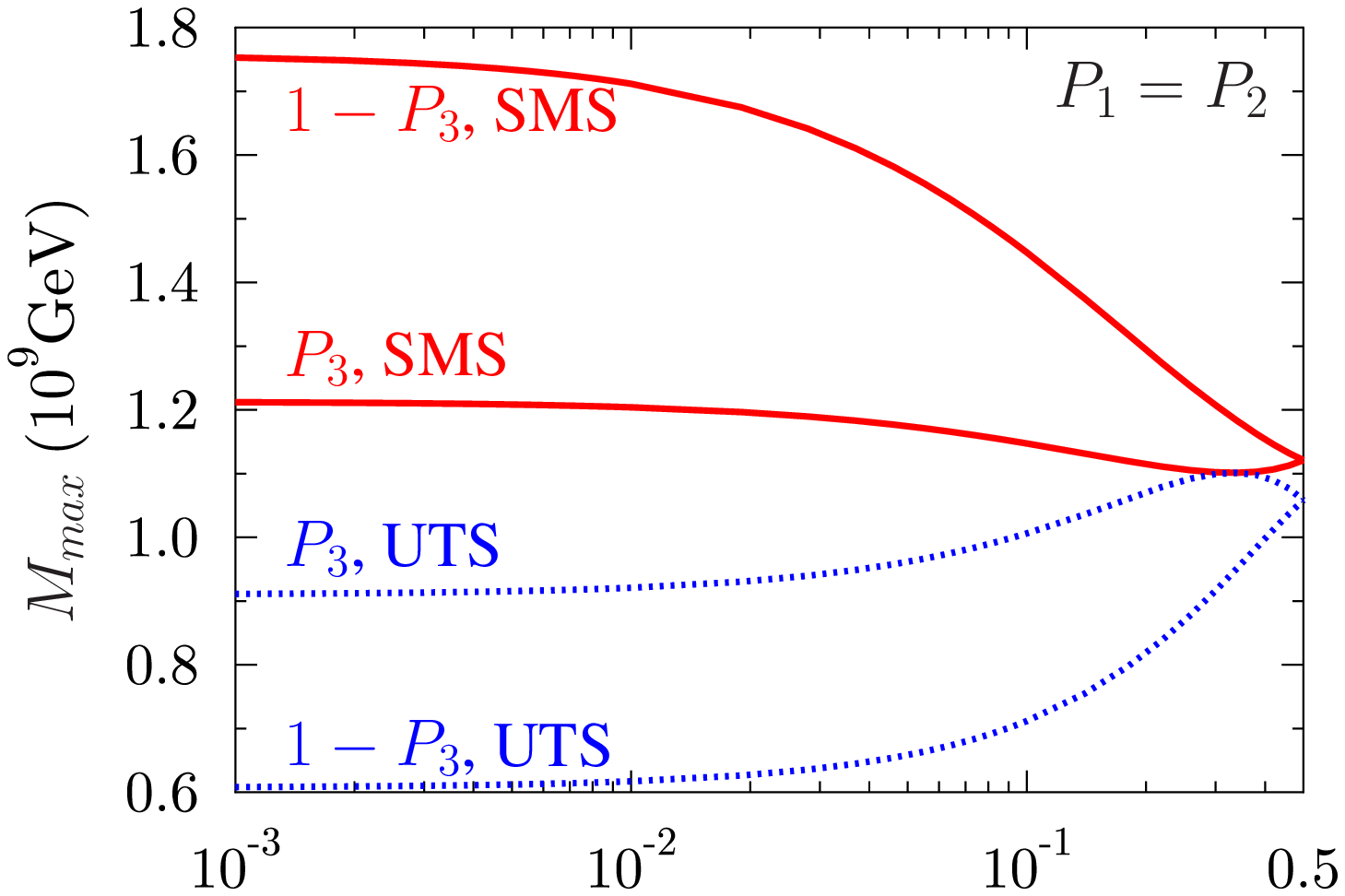,width=0.5\textwidth,angle=0}
\caption{Maximum values of $B$ and $M$ which 
can lead to successful leptogenesis 
as a function of the flavour projections
(we plot both as a function of $P_3$ or $1-P_3$ for clarity when either
$P_3$ or $1-P_3$ is very small).  
The figure corresponds to $A\sin\phi_A$=1 TeV and $\tan\beta=30$.} 
\label{fig:MmaxPBmaxP}
\end{figure} 

We next explore the impact that flavour enhancements can have in
relaxing the requirements on the values of $B$ and $M$ for successful
leptogenesis.  From
Eqs.~\eqref{eq:yb-l},~\eqref{eq:yb},~\eqref{eq:CP_asym},
and~\eqref{eq:wmapeta} we find that the maximum value of $B$ for 
given values of $M$ and $m_{\rm eff}$ is:
\begin{eqnarray}
B & \leq & \frac{\Gamma\left(m_{\rm eff},M\right)}{2}
\frac{\left|\mbox{Im}A\right|}{M}\frac{C\,\eta(m_{\rm eff})}
{Y_{B_{obs}}}\left[1+\sqrt{1-\left(\frac{M}
{\left|\mbox{Im}A\right|}\frac{Y_{B_{obs}}}
{C\eta(m_{\rm eff})}\right)^{2}}\right]\;,
\label{eq:B_pm}
\end{eqnarray}
where $C=\frac{16}{23}Y_{\widetilde{N}}^{eq}(z\to 0)$,
$\Gamma(m_{\rm eff},M)$ is given in Eq.\eqref{eq:gamma} and 
${\rm Im}A= A\sin\phi_A$.  
Thus we obtain  
\begin{eqnarray}
\label{eq:max_M}
M & \leq & 
\frac{\left|\mbox{Im}A\right|C\,\eta(m_{\rm eff})
}{Y_{B_{obs}}}\; , \\
B & \leq & \frac{3\sqrt{3} m_{\rm eff}}{32\pi v^2} \left(
\frac{\left|\mbox{Im}A\right|C\eta(m_{\rm eff})}{Y_{B_{obs}}}
\right)^2\; , 
\end{eqnarray}
where $\eta(m_{\rm eff})\equiv\eta(m_{\rm eff},P_j,Z_j)$ and we have
neglected all residual dependence of $\eta$ on $M$.  As seen in the
right panel of Fig.~\ref{fig:eff_fla}, assuming the SMS and for
sufficiently hierarchical $P_j$, $\eta(m_{\rm eff})$ decreases first
very mildly with $m_{\rm eff}$ and -- once all the flavours have
reached the strong washout regime-- it decreases roughly as $\sim
m_{\rm eff}^{-2}$.  Thus the product $m_{\rm eff}\times \eta(m_{\rm
  eff})^2$ first grows with $m_{\rm eff}$ till it reaches a maximum
and then for sufficiently large $m_{\rm eff}$ it decreases $\sim
m_{\rm eff}^{-3}$.  Therefore, for a fixed value of the projectors, 
the upper bound on $B$ does not
corresponds simply to the maximum allowed value of $m_{\rm eff}$, but
it has a more complicated dependence.

In Fig.~\ref{fig:MmaxPBmaxP} we show the maximum values of $B$ and $M$
obtained for both the UTS and SMS cases as a function of the flavour
projections. In order to have better resolution when either $P_3$ or
$1-P_3$ is very small, we plot them both as a function of $P_3$ or
$1-P_3$.  In the figure we set ${\rm Im}A=1$ TeV.  The figure
illustrates that within the UTS, the parameter space for successful
leptogenesis is very little modified by departing from the flavour
equipartition case (that corresponds to the point where the UTS and SMS
curves join). On the contrary, in the SMS case we find that with
hierarchical flavour projections $1-P_3 \sim {\rm few}\,\times
10^{-3}$ successful soft-leptogenesis is allowed also with $B\sim{\cal
  O}(TeV)$, that is for quite natural values of the bilinear term.  As
mentioned above, even for hierarchical projections the maximum allowed
values of $B$ and $M$ do not correspond to the maximum allowed value
of $m_{\rm eff}$.  In particular, for the range of flavour projections
shown in the figure we obtain that the maximum values of $B$ and $M$
correspond to  $m_{\rm eff}\lsim 2$ eV.

\begin{figure}[htb]
\centering
\includegraphics[width=0.8\textwidth]{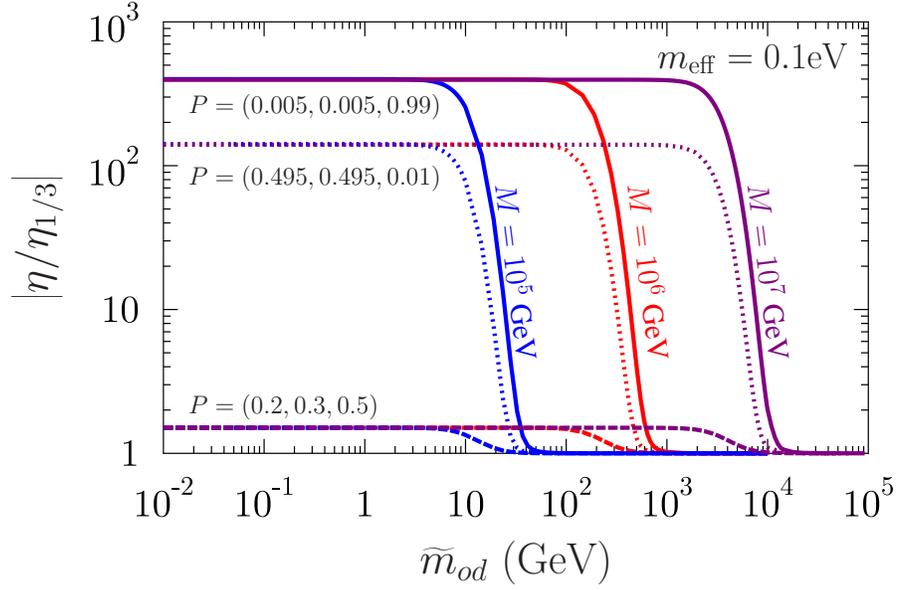}
\caption{The dependence of the efficiency (normalized to the flavour
  equipartition case $P=(1/3,1/3,1/3)$) on the off-diagonal soft
  slepton mass parameter $\widetilde{m}_{od}$, for different values of
$M$ and of the flavour projections (see text for details).}
\label{fig:eff_fci}
\end{figure}

We now turn to quantify the impact that the presence of LFE
scatterings can have on these results. We plot in
Fig.\ref{fig:eff_fci} the dependence of the enhancement of the
efficiency due to flavour effects, as a function of the off-diagonal
slepton mass parameter $\widetilde{m}_{od}$. As can be seen in the
figure (and as it was expected from the discussion in the previous
section) for any given value of $M$, LFE quickly becomes  efficient
damping completely the lepton flavours enhancements of the efficiency
within a very narrow range of values $\widetilde{m}_{od}^{min} \leq
\widetilde{m}_{od} \leq \widetilde{m}_{od}^{max}$. The figure is shown
for $\tan\beta=30$ . Again, the dominant dependence on $\tan\beta$ arises due to
$v_d=v\cos\beta$ in Eq.\eqref{eq:rij}. Results from other values of
$\tan\beta$ can be easily read from the figure by rescaling 
$\widetilde{m}^\beta_{od}=\widetilde{m}^{\rm fig}_{od}/(30 \cos\beta)$.

It is interesting to remark that while the efficiency $\eta(m_{\rm
  eff})$ is practically insensitive to the particular value of $M$, as
long as $m_{\rm eff}$ is held constant, this is not the case for the
generalized efficiency $\eta^{\rm LFE}$ computed by accounting for LFE
effects. Given the different scaling with the temperature of the
$\overline{\Gamma}_{\rm LFE}$ and $\overline{\Gamma}_{\rm ID}$ rates,
the precise temperature at which leptogenesis occurs is crucial.  For
example, we see from Fig.\ref{fig:eff_fci} that for reasonable values
$ \widetilde{m}_{od} \lsim 200\,$GeV and for $M\gsim 10^6\,$GeV,  LFE is
not effective, and  the large enhancements of the efficiency due
to flavour effects can survive, while for $M\lsim 10^5\,$GeV
all flavour enhancements disappear.

\begin{figure}[t!]
\centering
\includegraphics[width=0.8\textwidth]{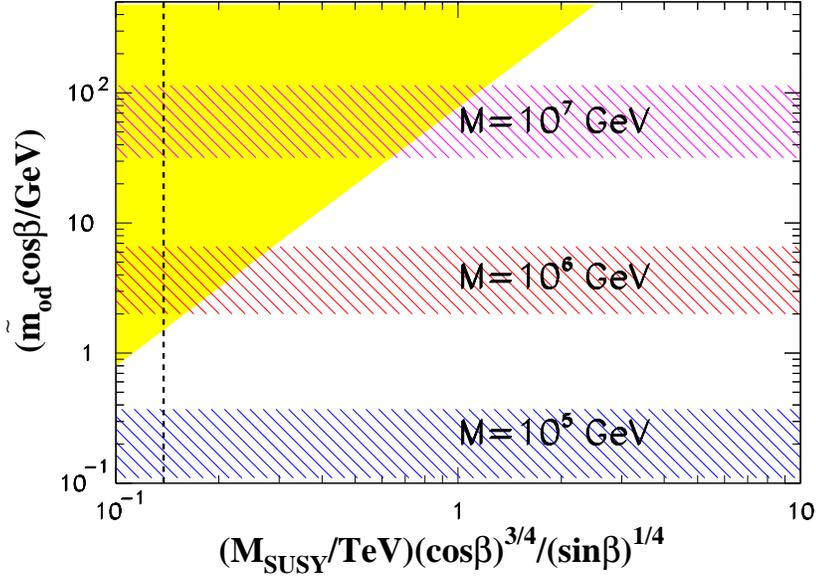}
\caption{Excluded region (shaded in yellow) of
  $\widetilde{m}_{od}\cos\beta$ versus
  $m_{SUSY}(\cos\beta)^\frac{3}{4}/(\sin\beta)^\frac{1}{4}$ arising
  from the present bound of $BR(\mu\rightarrow e\gamma)\leq 1.2 \times
  10^{-11}$, together with the minimum value of
  $\widetilde{m}_{od}{\cos\beta}$ for which LFE effects start damping
  out flavour effects in soft-leptogenesis.  Three bands are shown
  corresponding to $M=10^5\,$GeV, $M=10^6\,$GeV and $M=10^7\,$GeV.
  The width of the bands represents the range associated with
  variations of $P_k m_{eff}$ in the range $0.003\,$eV$-10\,$eV, where
  $P_k$ is the largest of the three flavour projections. The vertical dashed
  line represents the value of $m_{SUSY}/(\tan\beta)^\frac{1}{2}$ (for
  $\tan\beta=1$) required to explain the discrepancy between the SM
  prediction and the measured value of $a_\mu$~\cite{nuria}.}
\label{fig:susylfv}
\end{figure}

\section{Low energy constraints}
\label{sec:disc}

We have seen in the previous section that for not too large values $
\widetilde{m}_{od} \sim 200\,$GeV and for $M\gsim 10^6\,$GeV LFE effects
can be neglected.  In this section we address this issue more
quantitatively, comparing the values of $\widetilde{m}_{od}$ required
for total washout of flavour effects in soft leptogenesis, with the
bounds imposed from non-observation of flavour violation in leptonic
decays. The question we want to address is the following: given the low
energy constraints on $\widetilde{m}_{od}$, what is the lower bound on
the leptogenesis scale $M$ for which large flavour enhancements of the
lepton asymmetry are not damped by LFE effects ?

Clearly the presence of a sizable $\widetilde{m}_{od}$ would induce
various LFV decays, like for example $l_j \rightarrow l_k\gamma$
with rate
\begin{equation}
\frac{BR(l_j \rightarrow l_k\gamma)
}{BR(l_j
\rightarrow l_k \nu_j\overline\nu_k)}\sim
\frac{\alpha^3}{G_F^2}\frac{\tan^2\beta}{m^8_{SUSY}}
\widetilde{m}_{od}^4
\simeq 2.9\times 10^{-19} 
\frac{\sin^2\beta}{\cos^6\beta}\left(\frac{\rm TeV}{m_{SUSY}}\right)^8
\left(\cos^2\beta\frac{\widetilde{m}_{od}^2}{\rm GeV^2}\right)^2
\end{equation}
where $m_{SUSY}$ is a generic SUSY scale for the gauginos and sleptons
masses running in the LFV loop. We show in Fig.\ref{fig:susylfv} with
a yellow shade, the excluded region of $\widetilde{m}_{od}\,
{\cos\beta}$ versus
$m_{SUSY}(\cos\beta)^\frac{3}{4}/(\sin\beta)^\frac{1}{4}$ arising from
the present bound BR$(\mu\rightarrow e\gamma)\leq 1.2 \times
10^{-11}$, together with the minimum value of $\widetilde{m}_{od}\,
{\cos\beta}$ for which LFE effects start damping out flavour
enhancements in soft-leptogenesis.  Three bands are shown respectively
for $M=10^5\,$GeV, $M=10^6\,$GeV and $M=10^7\,$GeV.  The width of the
bands represents the range associated with variations of the effective
flavoured decay parameter $P_k m_{eff}$ in the range
$0.003\,$eV$-10\,$eV, where $P_k$ is the largest of the three flavour
projections. For illustration we also show in the figure the 
characteristic SUSY scale that allows to explain the small discrepancy
between the SM prediction and the measured value of
the muon anomalous magnetic moment, $a_\mu$.  This values is
$m_{SUSY}/(\tan\beta)^\frac{1}{2}=141$ GeV~\cite{nuria}, and the
vertical dashed line in the picture corresponds to $\tan\beta= 1$.  As
seen in the figure, in this case the off-diagonal slepton masses are
bound to be small enough to allow for flavour enhancements in soft
leptogenesis  for $M$ as low as $10^6$ GeV. For larger values of
$\tan\beta$, even lower values of $M$ are allowed.

\section{Discussion and Conclusions} 
\label{sec:concl}
Within the supersymmetric leptogenesis scenario, the generation of the
baryon asymmetry unavoidably receives contributions from dynamical
effects in the decays of the heavy sneutrinos that are specifically
related to soft SUSY breaking terms. The interesting point is that
soft leptogenesis effects become relevant at relatively low
temperatures, and precisely when the usual contributions surviving in
the limit of unbroken supersymmetry become ineffective to generate a
sufficient amount of lepton asymmetry. Soft leptogenesis thus opens up
a low temperature window where supersymmetric leptogenesis can proceed
without conflicting with the gravitino limits on the reheating
temperature. 

However, the efficiency of soft leptogenesis remains relatively low,
especially when flavour effects are neglected. This is mainly due to
the fact that the relevant CP asymmetries vanish in the zero
temperature limit. Some mechanism to generate sufficiently large
enhancements of the lepton asymmetry that is produced are then
required, and are generally obtained by assuming a resonant or
quasi-resonant regime for the decays of the pair of heavy sneutrinos
belonging to the same family. Rather unpleasantly, the resonant
conditions can be ensured only by requiring that the sneutrino mixing
parameter $B$, that is the parameter that controls the splitting
between the pairs of mass eigenvalues, has a value that is unnaturally
suppressed with respect to the SUSY breaking scale: $B\ll m_{SUSY}$.

However, given the temperature regimes in which soft leptogenesis can
proceed, accounting for flavour effects is mandatory. These effects
where first studied in ref.~\cite{oursoft} under the assumption of
universality of the soft SUSY breaking terms. Such scenarios strongly
constrain the possible flavour structures, and in particular imply
that the flavoured CP asymmetries must be proportional to the
corresponding flavour dependent washouts, with the result that the
larger is the CP asymmetry, the more efficient is the related
washout. This compensating mechanism allows for only moderate $\sim
{\cal O}(30)$ enhancements of the leptogenesis efficiency.  Thus,
within universal soft-breaking schemes, flavour effects can only
moderately alleviate the fine tuning problem of the $B$ parameter,
and still do not allow for $B \sim m_{SUSY}$, that is what one would
expect on the basis of naturalness considerations.

In this paper we have shown how this situation drastically changes if
the assumption of universality for the soft-breaking terms is relaxed,
which results in a generic situation in which the flavoured CP
asymmetries are not aligned with the respective washouts.  Note that
an analogous situation is generally realized within the standard
flavoured leptogenesis scenarios.  To carry out our phenomenological
analysis, while avoiding the proliferation of too many flavour-related
parameters, we have introduced a simplified non-universal scheme in
which all the flavoured CP asymmetries (evaluated at equal
temperatures) are equal (and thus flavour independent) while the
flavoured washouts are allowed to be strongly hierarchical. Here we
stress that since the hierarchy in the washouts is controlled by the
hierarchy in the sneutrino Yukawa couplings, and given that we know
that in the SM strong hierarchies in the Yukawa couplings are realized
in the charged lepton sector as well as for the up- and down-type
quark sectors, a strong hierarchy in the flavour dependent washouts
can be considered as a natural possibility.  As regards the amount of
misalignment between the soft-breaking trilinear terms and the
corresponding Yukawa couplings, that eventually produces the
misalignment between flavoured CP asymmetries and washouts, due to our
ignorance about the mechanism that breaks SUSY, any assumption is
equally acceptable, provided that the existing limits on LFV processes
are not violated.

Under these conditions, we have found that flavour effects can enhance
the leptogenesis efficiency by more than two orders of magnitude with
respect to the flavour equipartition case, defined as the situation in
which all the flavoured CP asymmetries and washouts are equal in
magnitude. This result can then be translated into a several$\,\times
10^3$ enhancement with respect to the one-flavour approximation, which
is sufficient to avoid the need for any additional enhancement from
resonant conditions. Thus, the natural scale for the sneutrino mixing
parameter $B \sim m_{SUSY}$ is eventually allowed.  Curiously, the
possibility of such large enhancements is directly related to the
strong temperature dependence of the CP asymmetries: for the lepton
flavours that are most weakly washed out, and for which inverse-decays
go out of equilibrium first, the lepton asymmetry is generated at
larger temperatures, that is precisely where the CP asymmetry is
larger.  Thus, relying only on the assumption of flavour misalignment
and of hierarchical Yukawa couplings, a very efficient scheme in which
the weaker is the washout, the larger is the corresponding CP
asymmetry, is automatically realized, and this boosts the
leptogenesis efficiency to the highest possible values.

\vspace{2truecm}

\acknowledgments 

This work is supported by  USA-NSF grant PHY-0653342 and by 
Spanish  grants from MICINN 2007-66665-C02-01, the 
INFN-MICINN agreement program ACI2009-1038,  consolider-ingenio 2010 
program  CSD-2008-0037 and by CUR Generalitat de Catalunya grant 2009SGR502.
EN acknowledges hospitality from the 
high energy physics group at  the University of Antioquia, 
where large part of this work was carried out.

\end{document}